# The preparation and properties of polycrystalline $Bi_2O_2Se$ - pitfalls and difficulties with reproducibility and charge transport limiting parameters

Jan Zich[1,2,*], Tomáš Plecháček[1], Antonín Sojka[1], Petr Levinský[2], Jiří Navrátil[1], Pavlína Ruleová[1], Stanislav Šlang[1], Karel Knížek[2], Jiří Hejtmánek[2], Vojtěch Nečina[3], and Čestmír Drašar[1]

[1]*University of Pardubice, Faculty of Chemical Technology, Studentská 573, 53210 Pardubice, Czechia*
[2] *FZU - Institute of Physics of the Czech Academy of Sciences, Cukrovarnická 10/112, 162 00 Prague 6, Czechia*
[3] *UCT Prague, Faculty of Chemical Technology, Technická 5, 166 28 Prague 6, Czechia*

*Corresponding author: jan.zich@student.upce.cz

## Abstract

Thermoelectric materials allow the direct conversion of waste heat into electricity, and novel materials are being investigated for this purpose. Recently, doped $Bi_2O_2Se$ has shown high application potential. In this study, we discuss causes for large variation in reported transport properties of pure $Bi_2O_2Se$ and present a preparation method that improves the reproducibility of undoped polycrystalline samples and improves their stability under thermal cycling. Key steps of this method include calcination of the $Bi_2O_3$ precursor, purification of the synthesized material in a temperature gradient, use of a coarse particle fraction and compaction of the powders in a $Si_3N_4$ die instead of a graphite die. The resulting polycrystalline material exhibits improved reproducibility and enhanced resistance to thermal cycling. It has room temperature electrical conductivity $\sigma_{RT} \approx 500$ S·m$^{-1}$ and Seebeck coefficient $S \approx -300$ µV·K$^{-1}$. These properties make it suitable as a reference material for future doping studies. The presented synthesis approach may provide a more reliable platform for investigating the intrinsic behavior and doping response of $Bi_2O_2Se$ in thermoelectric applications.

## Keywords

$Bi_2O_2Se$, polycrystals, transport properties, reproducibility, thermal cycling stability

# 1  Introduction

The search for novel low-cost and non-toxic thermoelectric materials with sufficient conversion efficiency is the primary focus of thermoelectric materials science today. $Bi_2O_2Se$ has become a widely studied quasi-2D semiconductor with promising transport properties of its single crystals, such as exceptionally high electron mobility, and for the ease of formation of $Bi_2SeO_5$ insulating layers, which is an important attribute in the fabrication and integration of electronic devices [1,2]. According to the literature, $Bi_2O_2Se$ in single crystal form is a well-characterized compound [3]. As a single-crystalline 2D material, it has been established as a photochemical detector [4], field-effect transistor [5] or oxygen detector [6]. In its polycrystalline form, $Bi_2O_2Se$ has been explored as a potential cost-effective, stable and easy-to-prepare alternative to tellurium-based materials in flexible and wearable thermoelectric electronics [7–9]. However, the poor reproducibility of reported properties in $Bi_2O_2Se$—particularly in the polycrystalline form—remains a significant obstacle. Inconsistencies in the reported properties of supposedly identical pure materials have been the motivation for this work.

Over the past decade, most of the research focused on the optimization of thermoelectric parameters through doping, namely electrical conductivity, Seebeck coefficient, and thermal conductivity [10–27]. A primary motivation for doping was to improve the low electrical conductivity of pure $Bi_2O_2Se$. However, doping frequently induces the formation of foreign phases (FPs), which complicates the material's stability and performance. This problem led to an alternative approach, which involves doping by the intentional addition of FPs to the pure $Bi_2O_2Se$ phase. In this context, several composite designs have been proposed [15,16,28,29]. Additionally, studies have explored the correlation between the non-stoichiometry of individual elements and the material's transport and thermoelectric properties [30–33]. The motivation for stoichiometric variation was to induce the formation of corresponding native defects. However, since the resulting concentration of native defects and thus the non-stoichiometry is much lower than expected, these materials often contain compositionally related phases such as $Bi_2O_3$, $Bi_2Se_3$, and $Bi_2SeO_5$.

Overall, the polycrystalline materials in the mentioned studies often lack phase purity, complicating the investigation of the intrinsic properties of polycrystalline $Bi_2O_2Se$. While X-ray diffraction is typically used to detect FPs, the identification of very thin 2D inclusions within the 2D matrix of $Bi_2O_2Se$ is challenging due to their low concentration, small size, and limited coherence length (Electronic Supplementary Material (ESM), section 6). This leads to the potential misinterpretation of the results as if they were for a phase pure material.

In addition, our review of the published data has revealed some significant inconsistencies:

1) There are considerable variations in electrical conductivity in undoped polycrystalline materials, with values for pure $Bi_2O_2Se$ ranging over 4 orders of magnitude in published studies. While most undoped samples exhibit metallic behavior [10,12,15–19,21,23,24], some exhibit semiconducting behavior [17,28,34,35]. This wide range in reported electrical conductivity can be attributed to variations in stoichiometry and the presence of FPs (Figure 1). The significant effect of non-stoichiometry on electrical conductivity and Seebeck coefficient is further shown in ESM Figure S2. As the number of FPs can increase substantially with doping (described below and in Table S6), it is challenging to distinguish the effects of doping from those due to variations in stoichiometry and the presence of FPs in doped samples. The problem of FPs complicating the interpretation of results was first highlighted in 2018 [24].

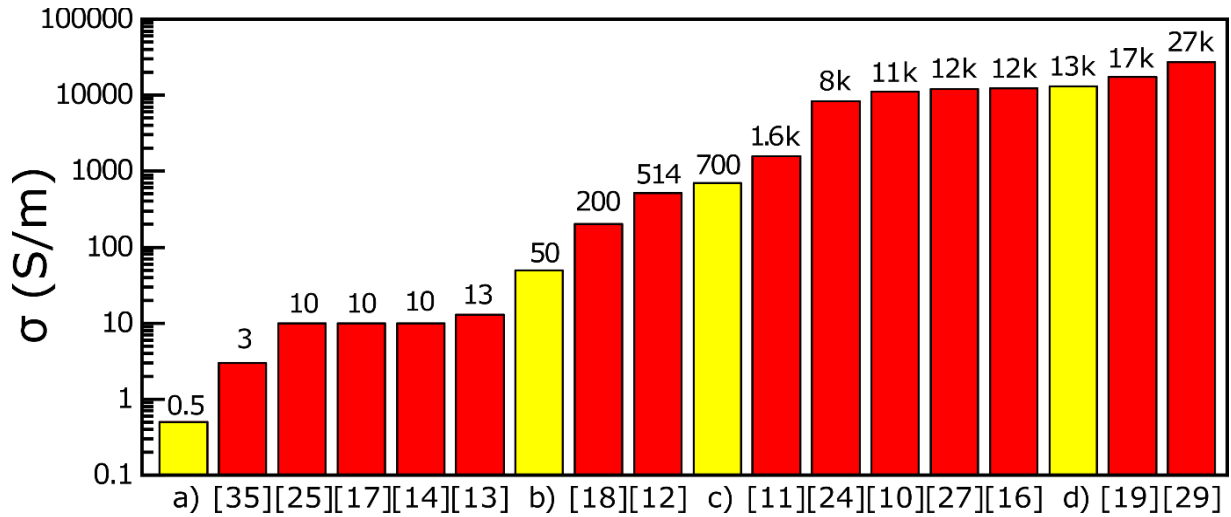

**Figure 1.** Comparison of the electrical conductivity $\sigma$ of undoped polycrystalline materials in various doping studies. Red columns represent literature data of undoped materials prepared by solid-state reactions at high temperature. Yellow columns represent our measured data from a batch of synthesized $Bi_2O_2Se$ with presented low-temperature method hot-pressed at 730°C, 70 MPa (a – < 35 µm fraction, $Si_3N_4$ die; b – all fractions, $Si_3N_4$ die; c – 35-340 µm fraction, $Si_3N_4$ die; d – all fractions, graphite die). Numbers above the columns are approximate values of room temperature electrical conductivity (S·m$^{-1}$), numbers below the columns are citation numbers. The preparation methods used in the cited studies are listed in Table S7. The 'k' in labels denotes ×10$^3$. The electrical conductivities of SPS prepared samples are given in ESM (Figure S7, S8) with explanation.

2) There is a lack of data in the literature on the thermal stability of both undoped and doped $Bi_2O_2Se$ under thermal cycling. Exploring this is crucial for its potential applications at elevated temperatures. In our experiments, polycrystalline samples of $Bi_2O_2Se$ showed notable changes in transport properties when subjected to thermal cycling (see Figures 5 and 6). This is because polycrystalline samples are composed of numerous small single crystals, where both the influence of grain boundaries combined with FPs and the intrinsic quality of each individual grain—primarily influenced by native point defects—all play significant roles.

The aim of this study is to describe the changes in the properties of undoped $Bi_2O_2Se$ based on synthesis/compaction approach. Through systematic analysis, we identify several important factors including precursor contamination, reaction with ampoule material, surface-driven phase formations, grain size effect, and pressing-induced reduction. All these factors contribute to the poor reproducibility observed across literature. Although we outline a processing strategy that partly improves the thermal and structural stability of $Bi_2O_2Se$, our aim was to highlight which synthesis variables are most critical and which remain unpredictable or uncontrolled, as shown in Figure 2. We further argue that even under tightly managed conditions, the material continues to exhibit complex behavior under thermal cycling, indicating that Bi$_2$O$_2$Se may be intrinsically prone to stoichiometric drift and grain boundary evolution. Our finding highlights the need for a careful approach to results of doping studies due to the current lack of stable and reproducible baseline material

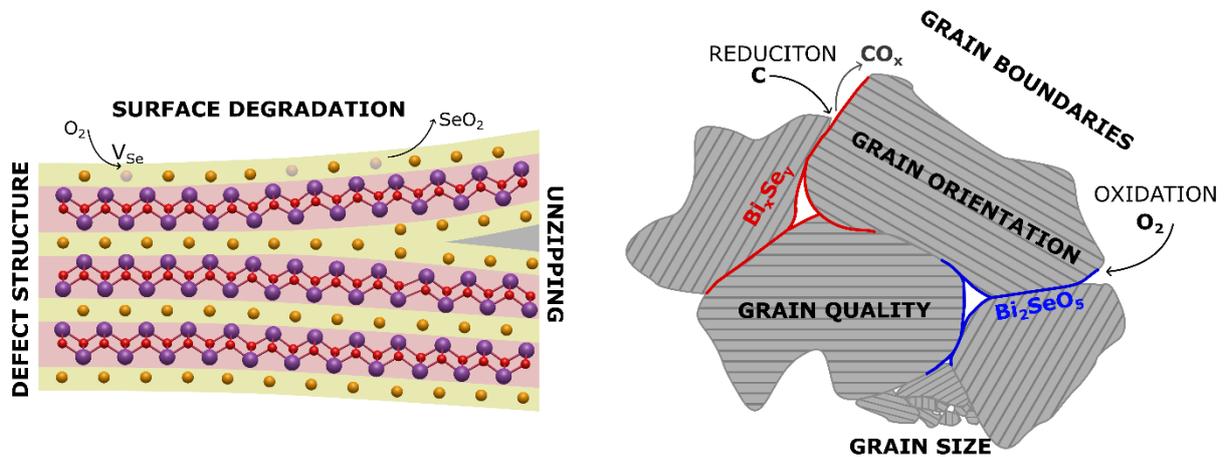

**Figure 2.** Schematics of Bi$_2$O$_2$Se crystal structure (left) and grains in polycrystalline sample (right) showing various procceses influencing properties of prepared polycrystalline samples.

## 2 Experimental

Bi$_2$O$_2$Se samples were prepared by both high and low-temperature synthesis. Bismuth chunks (SigmaAldrich, 5N) were ground in an MM500 nano oscillating mill at 30 Hz for 20 min, selenium shards (SigmaAldrich, 5N) were ground in an agate mortar, and Bi$_2$O$_3$ powder (AlfaAesar, 5N) was calcined (decarbonized) at 450 °C for 30 min in an argon flow and subsequently rapidly cooled to remove traces of Bi$_2$O$_2$CO$_3$. These powders were mixed in a stoichiometric ratio in an agate mortar and sealed in a quartz ampoule under a pressure of <10$^{-3}$ Pa.

The first set of samples was synthesized using a low temperature method, with a ramp of 1 °C/min to 400 °C, maintained for 10 days, followed by free cooling. To suppress the reaction between Bi$_2$O$_3$ and SiO$_2$, the surface of the ampoules was crystallized at 1000 °C for 48 h to improve chemical resistance and then cleaned with aqua regia, followed by deionized water, before use. To prevent agglomeration and ensure a homogeneous solid-state reaction, the ampoules were rotated in the tube furnace at approximately 1 rpm.

For comparison with synthesis methods from the literature, a second set of samples was synthesized using a common high-temperature method, with a ramp of 1 °C/min to 300 °C, maintained for 1 day, followed with a ramp of 1 °C/min to 700 °C, and maintained for 10 days, followed by free cooling. As in the low-temperature method, the ampoules were rotated in the furnace.

To further investigate the effect of particle size on the properties of polycrystalline $Bi_2O_2Se$, coarse grained powder was prepared. This was achieved by synthesizing $Bi_2O_2Se$ using the low-temperature method, followed by crystal growth in a temperature gradient of 860 °C to 800 °C maintained for 14 days. This process led to the formation of visible crystals, some of which reached sizes up to 1 cm. The crystals were sieved to obtain particle-size fractions for subsequent pressing.

The phase purity of the samples was checked by powder X-ray diffraction (PXRD) (Cu Kα, λ = 1.5418 Å) on a D8 ADVANCE DAVINCI diffractometer (Bruker AXS, Germany) with a Bragg-Brentano Ɵ-Ɵ goniometer (radius 280 mm) equipped with a LynxEye XE-T detector. Scanning was performed at room temperature from 5° to 90° (2Ɵ) in 0.01° steps with a counting time of 1.5 s (total step time 288 s).

The thermal properties of powdered $Bi_2O_2Se$ were studied with a STA 449 F5 Jupiter apparatus (Netzsch) operating in the DTA mode. Measurements were performed at a heating rate of 5 °C/min on powder samples (250 mg) placed in a platinum crucible with a silica insert and heated under a flowing $N_2$ atmosphere.

For compacting of powders, both spark plasma sintering (SPS) and hot press were used for comparison. Hot press (HP) samples were prepared using HP apparatus (HP 20-4560, Thermal Technology, USA). Exactly 2g of powder were poured into either a graphite die or a silicon nitride ($Si_3N_4$) die (MTI Corp., US) with an inner diameter of 12 mm under an Ar atmosphere. The temperature and pressure gradually increased for 45 min until reaching the target temperatures of 580 °C, 620 °C, or 730 °C, and pressure of 70 MPa. Temperature and pressure were maintained for 1 h for normal powders and 3 h for coarse fraction powders, followed by rapid pressure release. The samples were allowed to cool freely inside the press for approximately 3 h. The density of all the samples was calculated using the volumetric method. The density of the pressed tablets ranged from 95 to 100 % of the theoretical density with the median being 97 %.

Spark plasma sintering samples were prepared using SPS apparatus (HP D 10-SD, FCT Systeme, Germany). Exactly 5 g of powder were poured into graphite die with an inner diameter of 20 mm lined with graphite paper. The temperature was increased up to 580, 650 or 730 °C at 100 °C/min followed by dwell time of 5 min. Cooling was maintained at 50 °C/min. Pressure of 70 MPa was applied continually from 300 °C to the maximum temperature, then released within 1 min during cooling. All experiments were performed in vacuum. The temperature was measured by optical pyrometer in a non-through hole in top piston (5 mm from sample surface). The density of the pressed tablets ranged from 87 % to 95 % of the theoretical density.

The circular tablets obtained were cut into rectangular shapes (approximately 10 × 3 × 2 mm$^3$) for the measurement of transport properties using a custom-built, solvent-free diamond wire cutter. Final polishing was performed on a diamond grinding wheel to achieve parallel-sided rectangular samples.

The electrical conductivity σ was measured by the four-terminal method using an LSR-3 instrument (Linseis, Germany) over a temperature range from 300 K to 780 K on round or rectangular hot-pressed samples. The Seebeck coefficient S was measured using the static DC method on the LSR-3, with electrical conductivity measured simultaneously. The measurements were performed in a He atmosphere at an overpressure of 0.1 bar. The Seebeck coefficient was measured with an accuracy of ±7% and repeatability of ±3.5%. Electrical conductivity was measured with an accuracy of ±10% and a repeatability of ±5%. All the transport properties were measured perpendicular to the sintering pressing direction.

The microstructure analysis was performed using a scanning electron microscope (SEM) LYRA 3 GMH (Tescan, Czech Republic). Prior to the analysis, one set of samples was finely polished to 5000-grit finish, etched in 20% nitric acid for approximately 10 s and neutralized in a diluted ammonia solution. The second set of samples was left unpolished and unetched. Finally, the surfaces of all the samples were cleaned by brief sonication (≈10 s) in an ethanol and ammonia solution (80/20 by volume), rinsed in deionized waterand air-dried. The sample composition was analyzed with energy dispersive spectroscopy (EDS) Aztec X-Max 20 system at 5 kV (Oxford Instruments, UK).

# 3  Results and discussion

## 3.1  Difficulties common to both undoped and doped Bi$_2$O$_2$Se

The commonly used high-temperature synthesis procedures involve several critical challenges that are likely major contributors to the difficulty in reproducibility of transport properties reported in the literature. The challenges outlined apply to both doped and undoped Bi$_2$O$_2$Se-based materials.

   a) Purity of precursors and their reactivity with SiO$_2$ during synthesis [24,36,37]
   b) Air stability of as-synthesized Bi$_2$O$_2$Se powder [6]
   c) Reaction with die during compacting
   d) Equilibrium reactions between matrix and foreign phases – grain boundaries – stoichiometry

Other related problems arise in the context of doping. These are beyond the scope of this article but are addressed to some extent in Section 3.2.

   a)    Purity of precursors and their reactivity with SiO$_2$ during synthesis

Commercial high purity (99.999 %) $Bi_2O_3$ stored under ambient conditions often contains significant amounts of impurities, reducing the stoichiometric Bi content. The most significant impurity, $Bi_2O_2CO_3$, forms readily in air at room temperature [37], leads to a shift in stoichiometry of precursors if not accounted for. Calcination (decarbonation) at elevated temperatures to eliminate carbonate content is commonly noted in $Bi_2O_3$ studies but is mostly omitted in $Bi_2O_2Se$ studies. The reported decomposition temperature of $Bi_2O_2CO_3$ to β- and α- $Bi_2O_3$ varies significantly, but a temperature of 450 °C is generally sufficient [36]. To minimize the amount of impurities, $Bi_2O_3$ calcination is recommended immediately prior to the synthesis of $Bi_2O_2Se$ (see Experimental section) [40].

Polycrystalline $Bi_2O_2Se$ is commonly synthesized from elements or binary precursors by solid-phase reactions at high temperatures from 600 °C to 800 °C. At these temperatures, finding a chemically inert material that can fully withstand this specific combination of compounds is challenging. $Bi_2O_3$ exhibits strong corrosive properties toward most ceramic and glass materials, while selenium and bismuth readily react with most metals that could be used as a reaction vessel instead of quartz. To investigate the reaction between precursors (Se, Bi, and $Bi_2O_3$) and the quartz glass ampoule wall, a series of polycrystalline batches were synthesized at varying temperatures (400 °C, 730 °C, and 860 °C). The observed reaction can be summarized as:

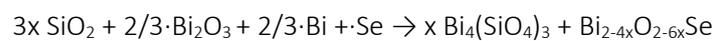

$$3x\ SiO_2 + 2/3 \cdot Bi_2O_3 + 2/3 \cdot Bi + \cdot Se \rightarrow x\ Bi_4(SiO_4)_3 + Bi_{2-4x}O_{2-6x}Se$$

As can be seen in Figure 3, the $Bi_4(SiO_4)_3$ covers the inner wall of the quartz ampoule. This reaction shifts the stoichiometry towards Se-rich, O- and Bi-poor composition. This conclusion is also supported by the presence of both $Bi_4(SiO_4)_3$ and $Bi_2Se_3$ detected by XRD on the inner wall of the ampoule. Interestingly, most of the synthesized powders show a single-phase diffraction pattern.

The reaction between $Bi_2O_3$ and quartz glass can be reduced by the surface crystallization of quartz through high-temperature annealing (1000 °C for 48 h). A thin layer of crystalline quartz exhibits higher resistance to reaction with $Bi_2O_3$, leading to the slower formation of $Bi_4(SiO_4)_3$. Furthermore, the reaction of precursors with quartz glass is much faster when these precursors are in the molten state, which limits the optimal synthesis to a solid-phase reaction [38,39]. Due to the mentioned reactions, a temperature around 400 °C with a prolonged synthesis time and well-mixed powders (to avoid the presence of unreacted $Bi_2O_3$ and $Bi_2Se_3$ in the product) is a preferred synthesis method.

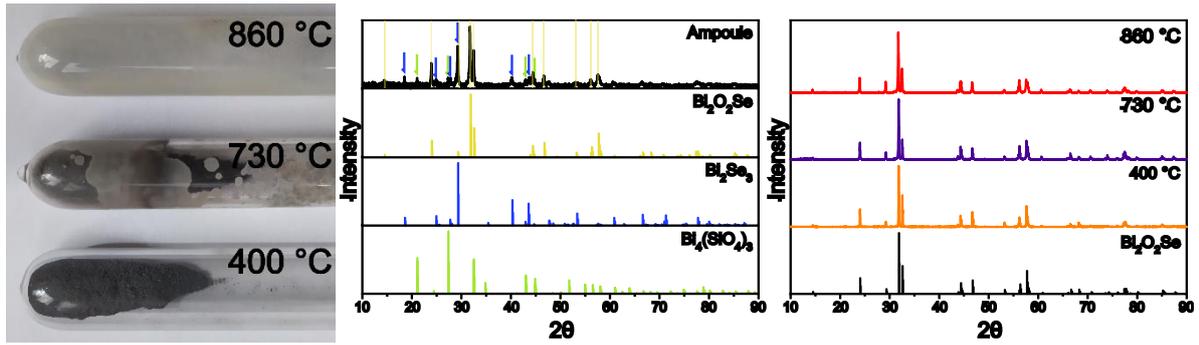

**Figure 3.** Reaction of Bi$_2$O$_2$Se with SiO$_2$ ampoules. The photo on the left shows the temperature dependence of the reaction rate of Bi$_2$O$_3$ with SiO$_2$ at 400 °C, 730 °C, and 860 °C. In the middle, the XRD pattern of the inner wall of the ampoules shows Bi$_2$SeO$_5$ formed on the ampoule surface. The right XRD pattern shows how products prepared at various temperatures can appear to be XRD pure independently of the reaction with the ampoules and adequately shifted stoichiometry.

b) Air stability of as-synthesized Bi$_2$O$_2$Se powder

During our experiments, as-synthesized pure polycrystalline Bi$_2$O$_2$Se powder handled and stored exclusively in a vacuum or inert argon atmosphere prior to hot pressing in a Si$_3$N$_4$ die could not be effectively densified, even at elevated temperatures (730 °C, 3 h, 70 MPa). The resulting pellets often failed to densify entirely or had lower density (< 95 %) compared to those sintered from the identical batches stored in ambient air. This difference is probably caused by adsorption of atmospheric oxygen on the particle surfaces, resulting in the formation of an amorphous surface phase driven by the electrostatic dipoles on the mosaic surface [6]. Notably, the estimated average adsorption energy (≈ 3.7 eV per O$_2$ molecule) is very high and comparable to the reaction enthalpies of common chemical reactions, such as carbon combustion, rather than adsorption energy. This issue was not observed during pressing in graphite die (see section c below).

As the surface Se layer vacancies fills with oxygen, the properties of the bulk material beneath it change, leading to a measurable decrease in electrical conductivity. This extreme sensitivity of Bi$_2$O$_2$Se nanoplatelets to the ppm level of O$_2$ has been used for Bi$_2$O$_2$Se ppm oxygen detectors [6]. Oxidation is reported on very thin (10-100 nm) and small (≈20 × 20 μm$^2$) single crystals but is difficult to study in bulk polycrystalline materials. In these materials, the amount of surface-adsorbed oxygen varies depending on the duration of exposure, temperature, and particle morphology and size - factors that are difficult to control simultaneously.

The data and the findings in ref. 6 are inconsistent with the general perception of Bi$_2$O$_2$Se stability in the single crystal form [5,40]. For a very thin sample (≈10 nm), the resistivity increased by two orders of magnitude after 10 min of exposure to air. While the cationic layer of Bi$_2$O$_2$ is extremely stable, the anionic part (S, Se, Te) is easily replaceable by other structures, such as $Bi_2O_2(OH)(NO_3)$ [41], $Bi_2O_2NCN$ [42] or similar structures like $Bi_2O_2CO_3$ [43]. These surface reactions cause the formation of foreign phases at grain boundaries during compaction. This modifies the transport properties of polycrystalline $Bi_2O_2Se$ in several ways, such as through changes in the concentration of point defects, modulation doping, or simply through inter-grain carrier scattering, all depending on the nature of the foreign phase [44]. The use of coarse particle fractions improves reproducibility by reducing the surface-to-volume ratio and helping optimal grain orientation, as discussed in section d) below.

c) Reaction with die during compacting

Compacting synthesized powders in a graphite die is possible at lower temperatures (e.g. 580 °C) compared to dies made of other materials (e.g., $ZrO_2$ and $Si_3N_4$). This is because $Bi_2O_2Se$ is reduced by carbon at temperatures above ≈400 °C, as shown in Figures S3 and S4, forming a $Bi_xSe_y$ phase that melts at a lower temperature and binds $Bi_2O_2Se$ powder particles together. The reaction between graphite die and compacted powder can be summarized as:

$$3\ Bi_2O_2Se + 3\ C \rightarrow Bi_4Se_3 + 2\ Bi + 3\ CO_2\ (\Delta H = +110\ kJ\ (+1.14\ eV)/\text{reaction})\ [45].$$

This chemical reduction is endothermic and takes place only at higher temperatures. However, the diffusion of $CO_2$ out of the press chamber shifts the reaction equilibrium toward the products.

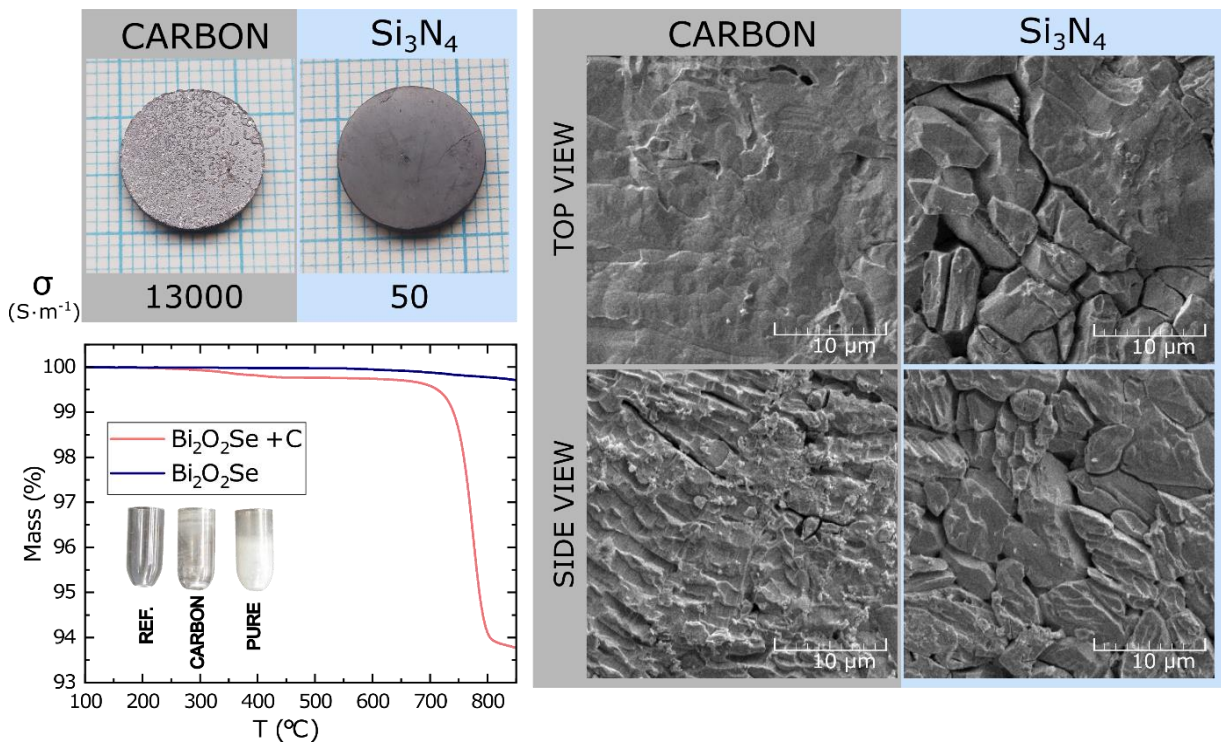

**Figure 4.** Comparison of optical images, SEM micrographs, thermogravimetry, and electrical conductivity of pellets hot-pressed at 730 °C in a C (graphite) and $Si_3N_4$ die. The SEM images of polished and etched samples pressed in a graphite die show increased $Bi_xSe_y$ content on the grain surfaces, that is responsible for an increase in electrical

conductivity. In contrast, no chemical reduction of $Bi_2O_2Se$ occurs in a $Si_3N_4$ die, resulting in mere particle compaction with no $Bi_xSe_y$ phase wetting the grain surface. This results in low electrical conductivity even in high-density samples. (A more detailed image with temperature dependence of C reduction is presented in Figure S4a). The onset temperatures of the ongoing reactions are evident in the thermogravimetric measurements.

In Figure 4, the DTA (TG) measurement shows a comparison of the reaction of pure sample and sample mixed with carbon in a stoichiometric ratio. In the pure sample, the reaction with the quartz crucible starts gradually around 500 °C, with the reaction rate increasing at higher temperatures. The slight mass loss observed can be attributed to the surface evaporation of selenium. The sample mixed with carbon begins to show a mass loss above 300 °C. This is caused by the surface reduction of the $Bi_2O_2Se$ particles and the formation of $Bi_xSe_y$ on the surface. This $Bi_xSe_y$ layer then acts as a protective barrier, shielding the interior of the particles from further reduction. Above 600 °C, the $Bi_xSe_y$ surface phase melts, allowing further reduction of the particle. With a stoichiometric ratio of $Bi_2O_2Se$:C of 1:1, a complete reduction to $Bi_xSe_y$ phases is observed. The total mass loss of 6% at 800 °C corresponds to all oxygen content in $Bi_2O_2Se$. It is worth noting that for the full reduction of a 2 g $Bi_2O_2Se$ (typical weight of compacted powder) to $Bi_xSe_y$, only 0.045 g of carbon is required. After hot-pressing the powder at 730 °C in a graphite die, small metallic Bi-Se spheres squeezed out of the die were observed, with the resulting pressed pellet reaching close to 100% theoretical density of $Bi_2O_2Se$ (Figure S4b and Table S2). This suggests that present foreign phases facilitates the compaction process of the $Bi_2O_2Se$ particles by acting as a lubricant/wetting agent at grain boundaries. $Bi_xSe_y$ phases were also observed on unpolished surfaces of SPS samples across all tested temperatures, with their content increasing with temperature.

For good grain contact, the mentioned $Bi_xSe_y$ phases appear to enhance intergranular charge transport [13,21,29,33]. However, this enhancement comes at the cost of altering the chemical identity of the sample and is also likely related to the instability of the transport properties during thermal cycling as suggested in section d). The presence of these secondary phases may also result in misleading conclusions regarding doping effects.

### d) Equilibrium reactions between matrix and foreign phases – grain boundaries – stoichiometry

The issues described in sections a) to c) likely result in a shift in stoichiometry and the formation of foreign phases, which may further react with the matrix during thermal cycling. The eventual change in composition depends on the temperature and its rate of change in terms of reaction kinetics and appears to be the key contributor to the phase instability of the bulk polycrystalline material under thermal cycling. While points a) and c) are directly solvable, the reaction with ambient oxygen discussed in section b) is difficult to address. During compaction at elevated temperatures, the oxidized surface of the powder kept in ambient conditions can also react with the particle core:

$$3\ Bi_2O_2Se + 2\ SeO_2 \rightleftharpoons 2\ Bi_2SeO_5 + Bi_2Se_3\ (\Delta H=-138\ kJ\ (-1.43\ eV)/reaction)\ [45],$$

$$2\ Bi_2O_2Se + 3\ SeO_2 \rightleftharpoons 2\ Bi_2SeO_5 + 3\ Se\ (\Delta H=-133\ kJ\ (-1.38\ eV)/reaction)\ [45].$$

If $CO_2$ and $SeO_2$ are present, the second reaction will apply, shifting equilibria in favor of the products:

$$3\ Bi_2O_2Se + SeO_2 + CO_2 \rightleftharpoons Bi_2SeO_5 + Bi_2Se_3 + Bi_2CO_5\ (\Delta H=-118\ kJ\ (-1.22\ eV)/reaction)\ [45].$$

Other possible reactions are listed in SI, section 5. These reactions are associated relatively low reaction enthalpies $|\Delta H|<0.4$ eV per $Bi_2O_2Se$ formula unit, suggesting that the equilibrium of these reactions can be significantly shifted with small changes in temperature or composition [46]. This tendency for surface driven reaction/decomposition appears to be a property of $Bi_2O_2Se$, inevitably leading to the formation of foreign phases at grain boundaries. Grain boundaries in conjunction with foreign phases play a critical role in the properties of polycrystalline $Bi_2O_2Se$ and its stability during thermal cycling (see Figures 5,6). One reasonable way of mitigating the effect of the particle surface/grain boundary is to reduce the total surface area of the particles used for compaction, thereby reducing the amount of oxygen adsorbed. This can be achieved by using coarse-grained, well-crystallized, low morphology material for compaction, that is, as-grown, uncrushed material.

The growth step in single crystal (SC) synthesis requires a considerable amount of time. In this regard, the surface crystallization of quartz is quite beneficial as it improves ampoule chemical resistance. Since the reaction of $Bi_2O_2Se$ with $SiO_2$ is slower than with $Bi_2O_3$, it is recommended to perform a low-temperature synthesis step before the high-temperature growth of SC. At high temperatures around 860 °C, the presence of a temperature gradient around 50 °C ensures the separation of the Se, $Bi_2Se_3$ and $Bi_2SeO_5$ phase on the cooler side of the ampoule, serving as a self-purification process. This results in the formation of pure well-crystallized $Bi_2O_2Se$ material on the warmer side of the gradient. The main problem is the formation of fine powder of $Bi_4(SiO_4)_3$ on the ampoule walls (see Figure S1). Due to the low vapor pressure, $Bi_4(SiO_4)_3$ is not transported to the colder side of the ampoule and remains with $Bi_2O_2Se$. However, it is a fine powder that can be largely removed by sieving (Table S1).

This high-temperature step produces both a few large single crystals and a significant amount of smaller crystals or crystal clusters, which can be effectively sieved into defined particle fractions suitable for compaction. The use of these well-crystallized coarse particle fractions minimizes surface for adsorption, leading to lower amount of foreign phases, and also promotes grain orientation during compaction (Figure 5). Samples prepared from these fractions exhibit higher reproducibility of transport properties and an improved resistance to thermal cycling (Figures 7, 8). Notably, for 2D materials, particle size selection by sieving is inherently suboptimal due to anisotropic shape and the frequent occurrence of intergrown crystallite clusters.

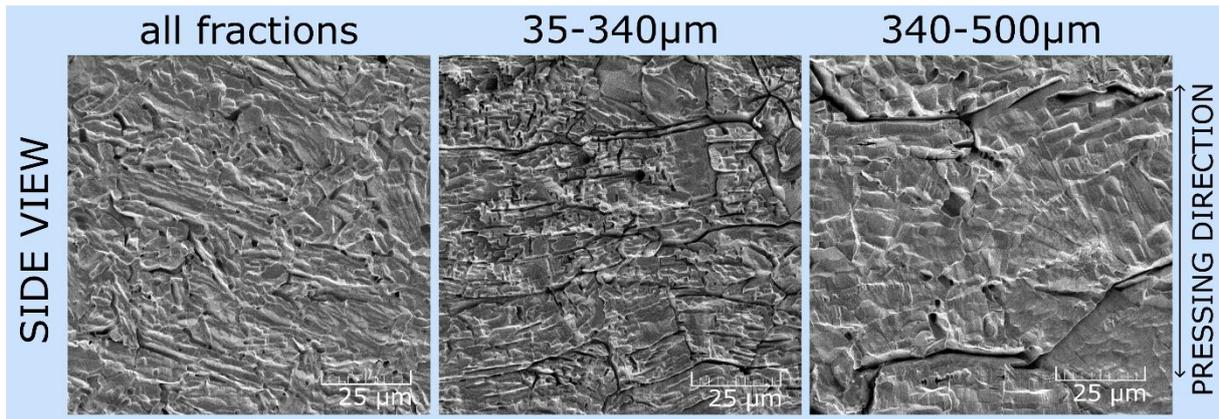

**Figure 5**. SEM images (100 × 100 μm$^2$) of samples pressed in a Si$_3$N$_4$ die from various particle size fractions. The images show that using a larger particle size fraction allows for better grain alignment within the sample, potentially improving the structural orientation of the material. In contrast, smaller particle sizes result in a more compact sample, which helps prevent the formation of intergranular voids. We also observed a correlation between particle size and electrical conductivity (Figure 1). Samples pressed from larger particles had a higher in-plane conductivity (perpendicular to the pressing direction), while those pressed from smaller particles had a lower conductivity. This reduction is likely due to the increased number of grain boundaries and suboptimal grain orientations.

## 3.2 Summary and results of improved synthesis method

To mitigate some of the most persistent synthesis-related problems, we propose a synthesis procedure that improves reproducibility of transport properties and enhances thermal cycling stability in polycrystalline Bi$_2$O$_2$Se. While these steps reduce certain sources of variability, they do not fully eliminate pure Bi$_2$O$_2$Se specific challenges.

**Low-temperature synthesis of pure Bi$_2$O$_2$Se**

1) Decarbonation of Bi$_2$O$_3$ at 450°C in Ar flow for 30 minutes prior to synthesis to minimize the content of Bi$_2$O$_2$CO$_3$ and thus maintain a correct stoichiometry.

2) Low-temperature synthesis of homogenized fine powder precursors at 400 °C in a crystallized SiO$_2$ ampoule to minimize the reaction of Bi$_2$O$_3$ with SiO$_2$ to Bi$_4$(SiO$_4$)$_3$ that also leads to stoichiometric change. Rotation of the ampoule during the synthesis likely accelerates the reaction. We assume this is due to gas phase reaction kinetics (this assumption is supported by how large single crystals grow).

**High-temperature gradient growth for optimization of grain boundaries**

3) Preparation of coarse particle fractions from pre-synthesized material in a crystallized SiO$_2$ ampoule at 860 °C in a temperature gradient. Secondary, the formed Bi$_2$SeO$_5$ and SeO$_2$ segregate in a cooler region of the ampoule and can be discarded, while Bi$_x$Se$_y$ phases remain in the hottest part of the ampoule walls (Figure S1).

4) Use of a defined coarse particle fraction for pressing, which reduces surface-to-volume ratio of the powder, minimizing the effect of ambient conditions and improving grain orientation in a pressed pellet. Traces of Bi$_4$(SiO$_4$)$_3$ formed during the high-temperature step have a small particle size and can be largely removed by sieving.

**Pressing**

5) The use of a Si$_3$N$_4$ die for pressing is essential to prevent the reduction of Bi$_2$O$_2$Se to Bi$_x$Se$_y$ phases. Pressing at temperatures close to the melting point (730 °C) is necessary to achieve a high density of samples.

While spark plasma sintering with tungsten carbide (WC) die might serve as a suitable alternative for graphite die, such setups are expensive and relatively uncommon. In our experiments with graphite dies, we attempted to prevent direct contact between the Bi$_2$O$_2$Se powder and graphite by using a separating layer of ZrO$_2$ powder, but this approach was only partially sucessful. The use of BN powder resulted in the formation of significant amount of BiSe phase and several unidentified peaks in PXRD.

By following the presented synthesis steps, we achieved improved reproducibility and stability in thermal cycling, as documented in Figures 7 and 8.

To evaluate the effects of individual synthesis parameters, we present 4 samples following some or all suggested steps. All samples were pressed in a Si$_3$N$_4$ die to prevent the formation of the Bi$_x$Se$_y$ phase.

**Sample 1** was prepared by a common synthesis method presented in the literature, with a synthesis temperature of 700°C for 10 days, without prior Bi$_2$O$_3$ calcination. No high-temperature growth in gradient or sieving was performed.

**Sample 2** followed low-temperature synthesis, but without high-temperature growth, temperature gradient purification or sieving.

**Sample 3 and 4** incorporated both low-temperature synthesis and subsequent high-temperature gradient growth. Only 35–340 µm particle fraction was used for pressing. In Sample 4, the powder was stored under ambient conditions for 1 month before pressing, to assess the influence of air exposure and surface oxidation.

All four samples showed a single-phase PXRD pattern of $Bi_2O_2Se$, although the presence of undetectable nanoscale or amorphous foreign phases cannot be excluded. The density of all the samples was > 98 % of the theoretical density.

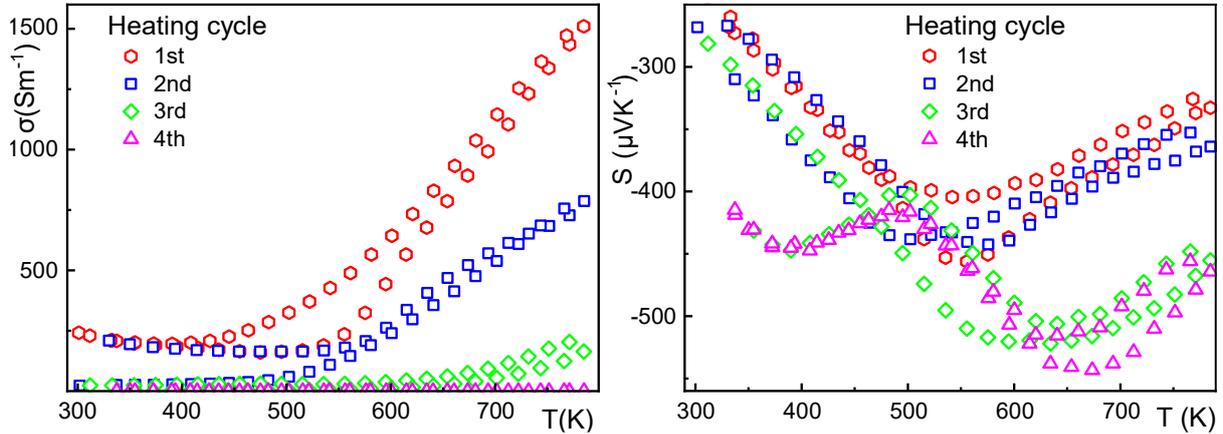

Figure 6. Sample 1 - electrical conductivity, Seebeck coefficient, and the effect of thermal cycling up to 500 °C. Electrical conductivity shows significant hysteresis during cycling, lowering electrical conductivity by two orders of magnitude with subsequent cycling. The Seebeck coefficient also shows hysteresis with a complex curve for evaluation.

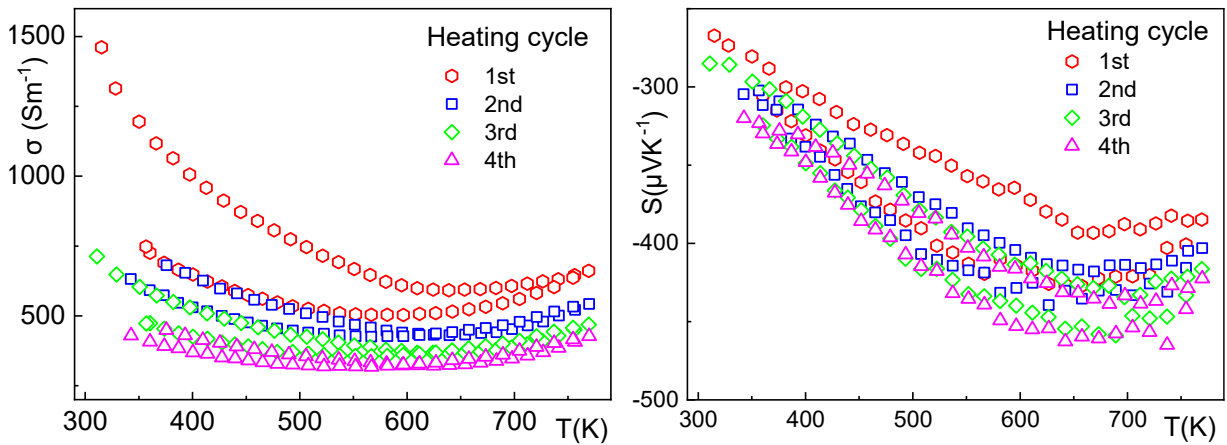

Figure 7. Sample 2 - electrical conductivity, Seebeck coefficient, and the effect of thermal cycling up to 500 °C. Electrical conductivity shows hysteresis during the first cycle, stabilizing after the third cycle around 500 S·m$^{-1}$ at 300 K. The Seebeck coefficient also shows slight hysteresis during the first cycle.

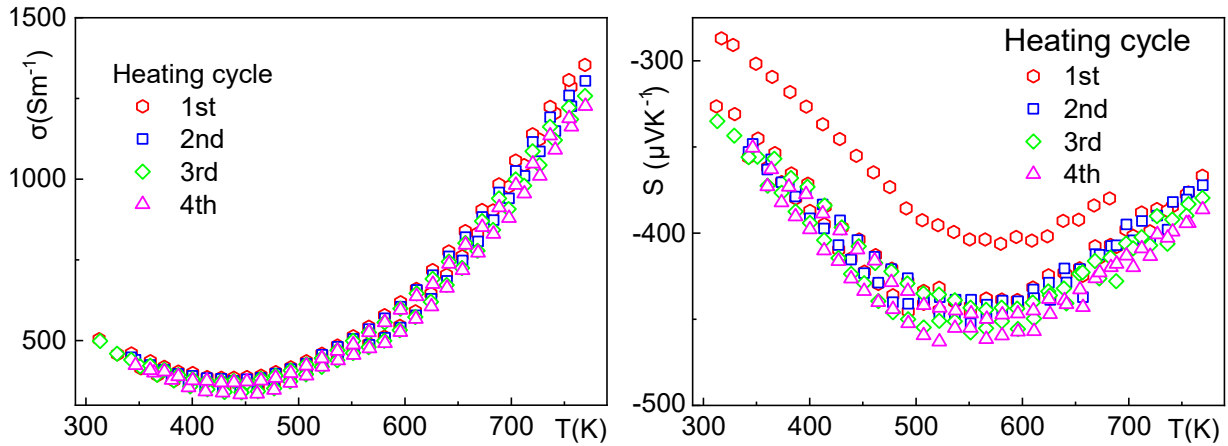

**Figure 8**. Sample 3 - electrical conductivity, Seebeck coefficient, and the effect of thermal cycling up to 500 °C. Electrical conductivity does not show any hysteresis with increasing the cycle count. Similarly, to sample 2, the Seebeck coefficient shows hysteresis during the first cycle, but stabilizes afterward.

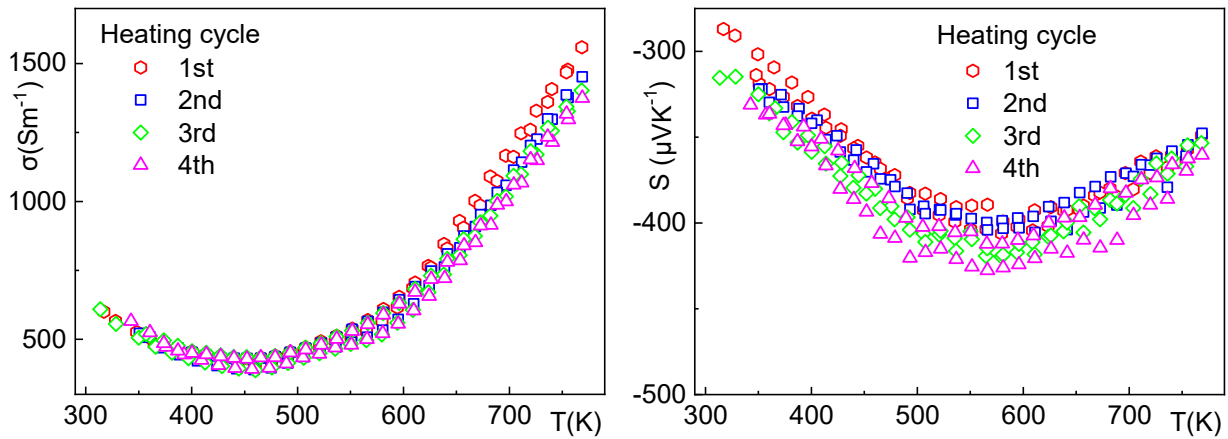

**Figure 9**. Sample 4 - electrical conductivity, Seebeck coefficient, and the effect of thermal cycling up to 500 °C. Both the electrical conductivity and Seebeck coefficient remain stable. A small decrease in electrical conductivity can be attributed to the slow surface oxidation and formation of an insulating layer of $Bi_2SeO_5$ due to the ppm concentration of $O_2$ in Ar used as an inert gas for measurement.

For samples 3 and 4 (Figures 8, 9), the electrical conductivity shows an exponential growth at temperatures above 500 K. The activation energy calculated from the temperature dependence with Arrhenius analysis is $E_A \cong 0.20$ eV, which would translate to an energy band gap of $E_g \cong 0.40$ eV for fundamental excitation. This is about half of the theoretical value of the band gap $E_g \cong 0.85$ eV reported in the literature, e.g. [47], indicating that the polycrystalline material may still deviate from ideal stoichiometry and likely contains native defects and/or foreign phases at grain boundaries. This is consistent with the slight decrease in electrical conductivity with cycling in sample 4. While the stability of the samples under thermal cycling has improved, we believe that the quality of the grain boundaries continue to influence transport behavior.

Additional cycling stability results, including comparative analysis of hot-pressed samples with various surface treatments and spark plasma sintered (SPS) samples, are presented in the ESM (Figures S5 – S8 and Tables S2 – S5). These data confirm that even samples exhibiting phase purity by PXRD may undergo significant changes in transport properties during cycling, depending on factors such as porosity, particle surface condition, or inhomogeneities introduced during pressing.

For further comparison, the properties of single crystals from this batch are shown in Figure S15. The band gap could not be measured from the temperature dependence of the electrical conductivity due to contact instability at temperatures above 150 °C. This highlights additional challenges related to the fabrication of long-term stable contacts for potential high-temperature thermoelectric applications.

### 3.3 Difficulties associated with doping of $Bi_2O_2Se$

Published data and our own experiments indicate that the solubility of the dopants in the $Bi_2O_2Se$ matrix is often very limited [24], Table S6. Ti-doped material [16] can serve as an example. In $Bi_{2-x}Ti_xO_2Se$, $TiO_2$ segregates as a foreign phase instead of substituting in the $Bi_2O_2$ layer. This leaves an O-poor and Se-rich composition of the matrix:

$$Bi_{2-x}Ti_xO_2Se \rightarrow Bi_{2-x}O_{2-2x}Se + xTiO_2.$$

This stoichiometric change has implications on the character and concentration of native defects and foreign phases (at grain boundaries). Since the Bi-O layer in $Bi_2O_2Se$ is extremely stable, doped materials focusing on Bi substitution often contain foreign phases. As a result, doping often leads to the formation of compositionally related foreign phases, e.g. BiSe, $Bi_4Se_3$, $Bi_2Se_3$, and dopant oxides. Among lanthanoids, Ce [17] and La [35] show reasonable solubility in the $Bi_2O_2Se$ matrix, as evidenced by the validity of Vegard's law. The effects of doping and connected stoichiometric changes are summarized in Figures S2a, b, ESM sections 4 - 6, and Table S6).

A better doping approach is the direct use of a specific foreign phase [15,16,28,29]. Such a phase can serve both as a facilitator of charge transport between grains and as a phase for modulation doping. The doping phase should ideally be stable in contact with $Bi_2O_2Se$ and should not form any potential barrier. This is still a challenge. In this direction, it is worth noting that the relative permittivity of $Bi_2O_2Se$ is high, close to $\varepsilon_r \approx 200$ [48], which is very favorable for modulation doping [44] and is unique for a non-ferroelectric material.

## 4 Conclusion

$Bi_2O_2Se$ is a unique material with ultra-high mobility, but it presents a challenge in achieving consistent charge and heat transport properties, especially in polycrystalline form. Our results confirm that polycrystalline samples exhibit limited reproducibility of the transport parameters, while these parameters also change due to temperature cycling. These problems are closely related to stoichiometry, leading to changes in the concentration of native point defects, the concentration of foreign phases, and the nature of grain boundaries, all of which remain difficult to control, even with careful synthesis procedure. We present several synthesis improvements that mitigate key sources of variability of polycrystalline $Bi_2O_2Se$, resulting in improved reproducibility and stability of the sample properties under temperature cycling. The presented measures include the calcination of the $Bi_2O_3$ precursor to maintain the correct stoichiometry, low-temperature synthesis of $Bi_2O_2Se$ in a crystallized silica ampoule, high-temperature crystal growth of the presynthesized $Bi_2O_2Se$ and purification of the material in a temperature gradient, using a defined particle fraction for pressing to aid particle orientation and minimize the number of foreign phases and grain boundaries, and finally pressing the powder in a $Si_3N_4$ die, rather than the conventional graphite die to prevent the reduction of $Bi_2O_2Se$ to $Bi_xSe_y$ phases. Implementation of these steps improves reproducibility and partially mitigates degradation effects during thermal cycling, though full property stabilization remains unresolved. At room temperature, the as-prepared pure sample pressed from the 35 – 340 µm particle fraction has an electrical conductivity of $\sigma_{RT} \approx 500$ S·m$^{-1}$ and a Seebeck coefficient of S $\approx$ -300 µV·K$^{-1}$ with a corresponding figure of merit $ZT_{RT}$ of 0.003 (Figure S16). This synthesis approach may serve as a useful reference for further systematic studies, particularly in the context of doping and microstructure–property correlations.

Acknowledgments

The authors would like to thank the Czech Science Foundation for financial support (Project No. 22 - 05919S), financial support from the grant of the Ministry of Education, Youth and Sports of the Czech Republic (grant LM2023037), and P. Mošner and T. Hostinský from the University Pardubice for the DTA measurements.

# Electronic Supplementary Material:

# The preparation and properties of polycrystalline Bi$_2$O$_2$Se- pitfalls and difficulties with reproducibility and charge transport limiting parameters

Jan Zich[1,2,*], Tomáš Plecháček[1], Antonín Sojka[1], Petr Levinský[2] Jiří Navrátil[1], Pavlína Ruleová[1], Stanislav Šlang[1], Karel Knížek[2], Jiří Hejtmánek[2], Vojtěch Nečina[3] and Čestmír Drašar[1]

[1]University of Pardubice, Faculty of Chemical Technology, Studentská 573, 53210 Pardubice, Czech Republic
[2] FZU - Institute of Physics of the Czech Academy of Sciences, Cukrovarnická 10/112, 162 00 Prague 6
[3] UCT Prague, Faculty of Chemical Technology, Technická 5, 166 28 Prague 6, Czech Republic

[*]Corresponding author: jan.zich@student.upce.cz

In the supplementary materials, we summarize some details that corroborate the conclusions drawn in the main text.

## 1. Purification of as synthesized material

For the preparation of coarse-grained material, we allowed the grains to grow in a temperature gradient and with the addition of bismuth selenite, Bi$_2$SeO$_5$. Growth in a temperature gradient has a self-cleaning effect. All foreign phases (Bi$_2$Se$_3$, Bi$_2$SeO$_5$, SeO$_2$) condense in the cooler part of the ampoule.

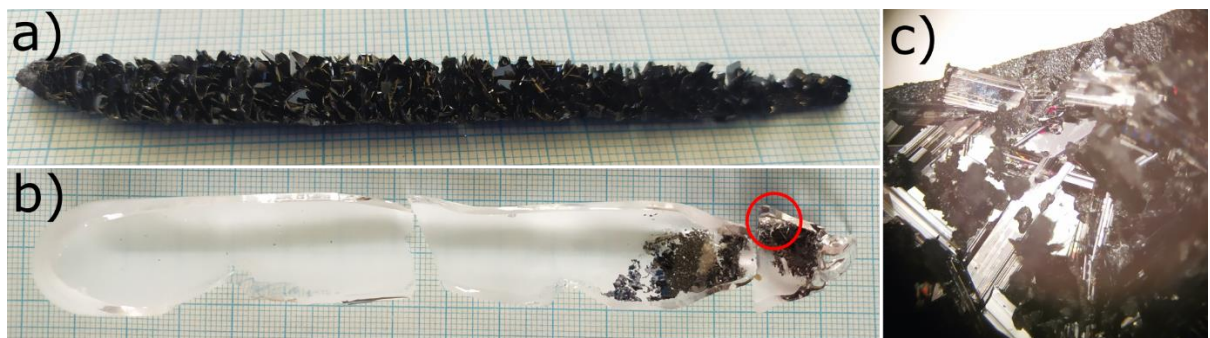

**Figure S1.** a) Single crystals of Bi$_2$O$_2$Se grown by physical transport in temperature gradient 800 °C/700 °C for 14 days, with left part (hotter side) contains Bi$_x$Se$_y$ phase and is discarded; b) Ampoule after synthesis/crystal growth with visible Bi$_2$SiO$_5$ formed on ampoule surface and Bi$_2$SeO$_5$ formed on cold side of temperature gradient; c) Focus on the Bi$_2$SeO$_5$ crystals from b) – red circle, indicating the presence of excess SeO$_2$, as discussed in points a) and c) in the main text. Most of the unwanted silicate phase is eventually removed by sieving (Table S1).

Care must be taken to maintain the correct gradient when cooling to room temperature. Such purification is insufficient in removing bismuth silicate Bi$_4$(SiO$_4$)$_3$. However, the use of a coarse fraction, i.e. careful sieving, helps to remove most of the fine-grained silicates from the material, as shown by the PXRD experiment (Table S1).

*Table S1*. Lattice parameters and foreign phases detected by PXRD of samples. The use of coarse-grained fractions helps to remove bismuth silicate, $Bi_4(SiO_4)_3$, which cannot be removed by heating in a temperature gradient (see main text).

| Sample | Lattice parameters | | | Main phase | Foreign phases |
|---|---|---|---|---|---|
| | a | c | V | $Bi_2O_2Se$ | $Bi_4(SiO_4)_3$ |
| $Bi_2O_2Se$ *(< 35 µm)* | 3.88591 | 12.2024 | 184.260 | 93.3 % | 6.7 % |
| $Bi_2O_2Se$ *(35 – 340 µm)* | 3.88579 | 12.2005 | 184.219 | 99.1 % | 0.9 % |

The reaction with $SiO_2$ is also problematic for preparation of single crystals in evacuated closed quartz ampoules. The shift of stoichiometry is accompanied by a shift in electrical conductivity as shown in Figure 1, Figure S2 and literature [1]. Moreover, diffusion of $Bi_2O_3$ into $SiO_2$ may be accompanied by diffusion of Si in the opposite direction, and $Bi_2SiO_5$ may be formed in the $Bi_2O_2Se$ volume [2]. In DTA measurement, we have observed onset of solid-solid reactions between $Bi_2O_3$ and $SiO_2$ around 500 °C, leading to formation of metastable $Bi_2SiO_5$. Temperature above 700 °C leads to formation of $Bi_4(SiO_4)_3$. Above 817 °C, the reaction between molten $Bi_2O_3$ and solid $SiO_2$ is very fast [3].

2. Difficulties with keeping the stoichiometry $Bi_2O_2Se$

We have prepared a series of $Bi_2O_2Se$ samples with deliberately altered stoichiometry; Bi- and O-stoichiometries vary from 1.98 to 2.02 and Se-stoichiometry varies from 0.98 to 1.02. Figures S2 show the electrical conductivity and Seebeck coefficient of polycrystalline $Bi_2O_2Se$ samples with the stoichiometry of two elements deliberately changed to reveal the correlation between the non-stoichiometry of the elements. These samples were prepared by a common method found in literature and hot pressed in graphite die at 730 °C for 1.5h. Their preparation route did not follow any recommendation for better stability given in the main text. The samples electrical conductivity vary by up to 4 orders of magnitude, which highlights effects of non-stoichiometry on sample properties. It is important to note that these changes in Bi:O:Se ratios can be unintentionally caused by substitutional doping in case dopant doesn't incorporate into $Bi_2O_2Se$ matrix.

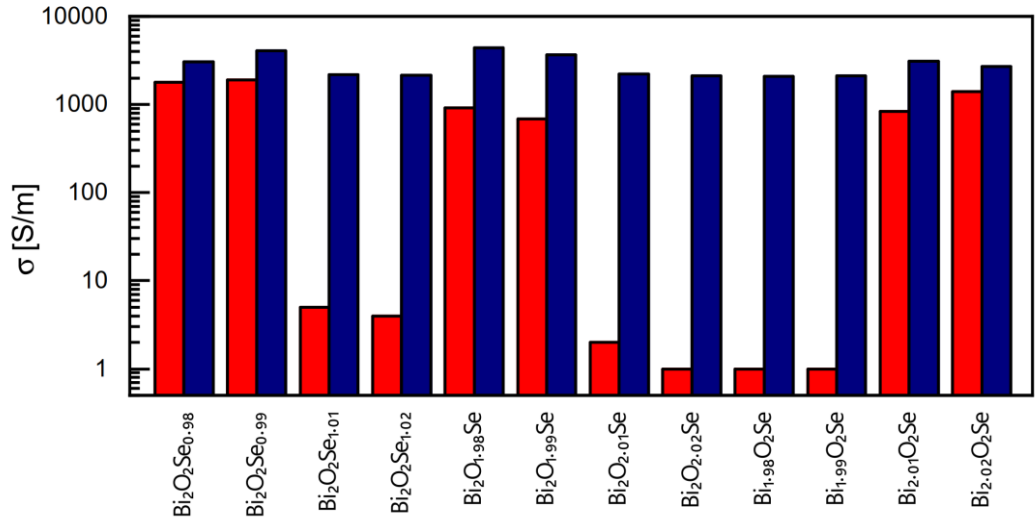

**Figure S2**. Bar graph showing electrical conductivity versus a specific non-stoichiometry for two temperatures, 300 K (red) and 723 K (blue) for first measurement cycle. The samples were prepared from crushed unsieved material was pressed in $Si_3N_4$ matrices for 1 hour at 730 °C.

Problems with maintaining stoichiometry also arise when pressing powders with a graphite die. Graphite reduces $Bi_2O_2Se$ to metallic elements and their compounds as documented in Figures S3 and S4.

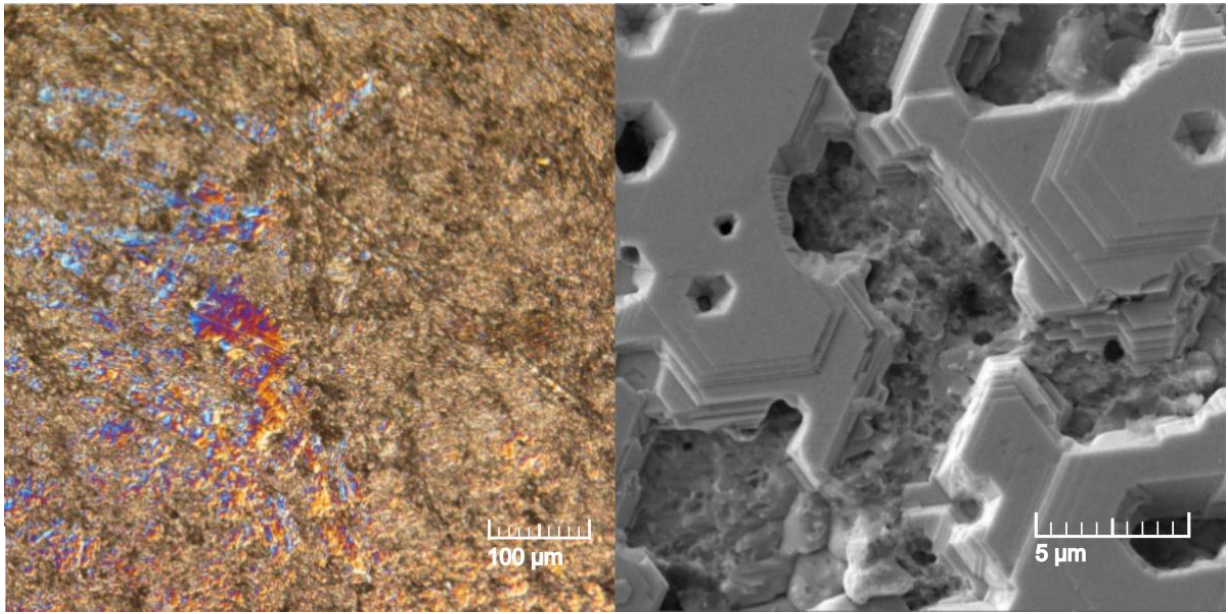

**Figure S3**. Polarized VIS and SEM images of $Bi_2O_2Se$ surface as pressed in graphite die at 580 °C. Polarized light on optical image (left) shows metallic parts, supported by SEM image (right) with visible hexagonal crystals of $Bi_xSe_y$ phases (mostly $Bi_2Se_3$ and $Bi_4Se_3$) growing on $Bi_2O_2Se$ surface (composition of phases was identified by EDS)

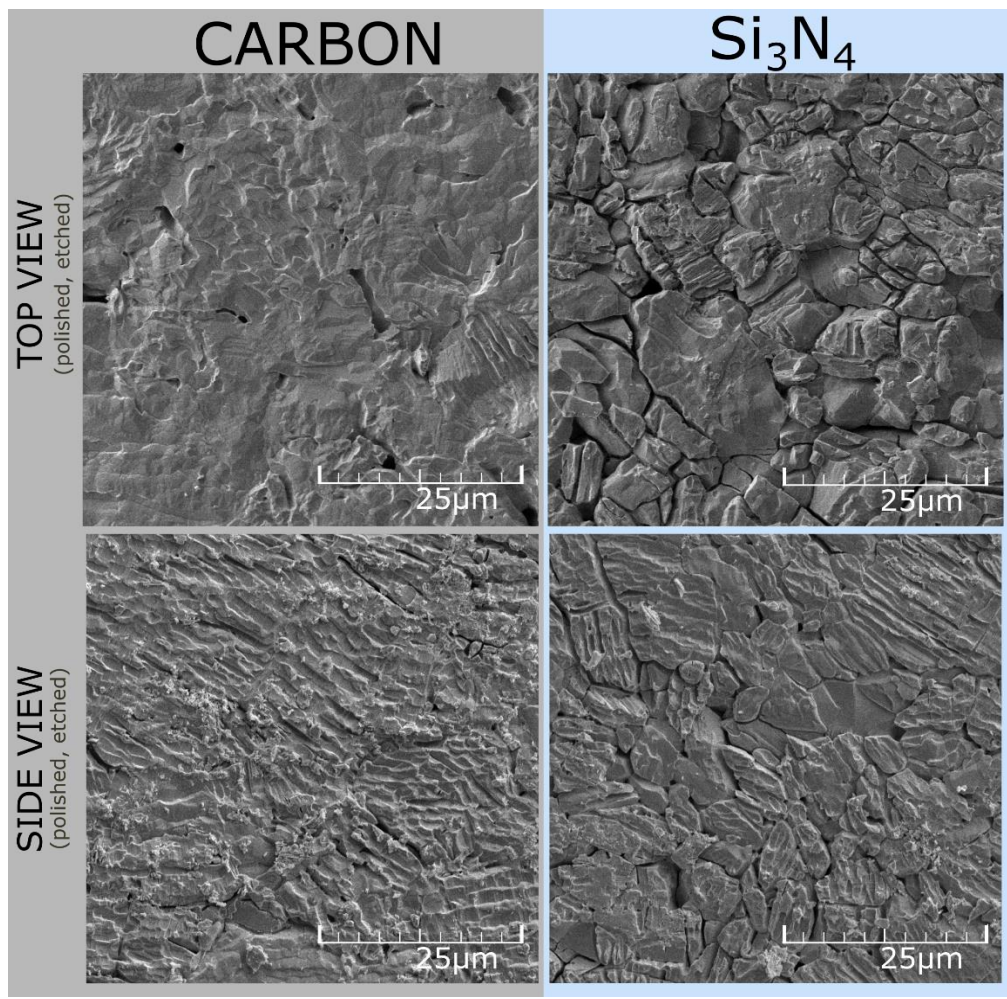

**Figure S4a**. Comparison of $Bi_2O_2Se$ etched surfaces pressed in graphite die and $Si_3N_4$ die at 730 °C. The graphite die sample shows better grain compaction and intergranular contact than the $Si_3N_4$ sample due to the presence of Bi-Se phases which etch significantly slower. Top view" refers to the top of the pellet and "side view" refers to the cross-section of the pellet.

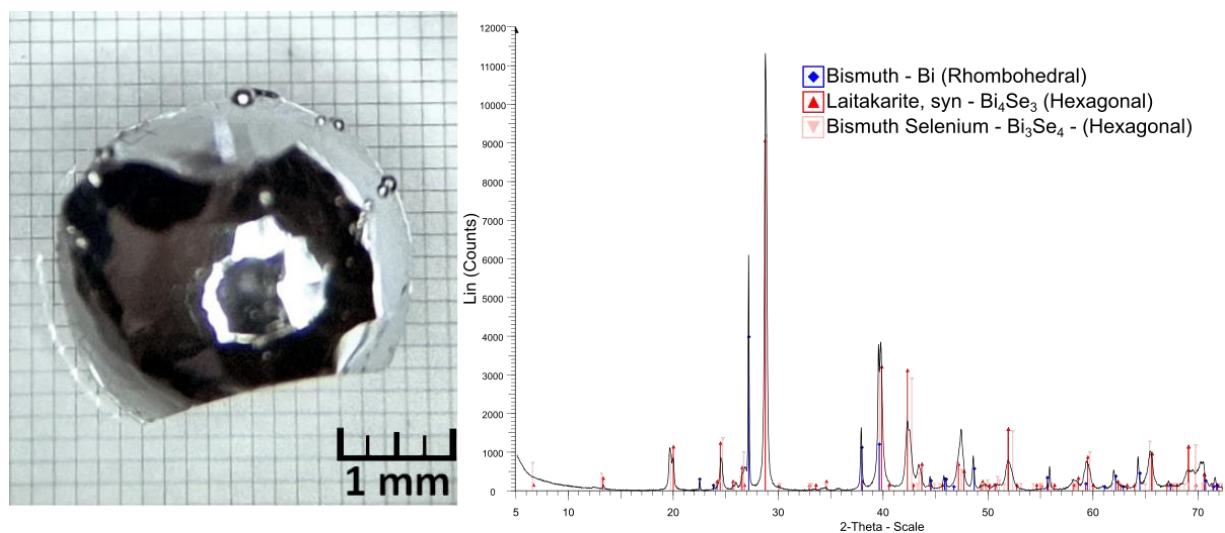

**Figure S4b**. Image of a flash from pressing in a graphite die at 730 °C and its PXRD analysis. The flash sphere consists of $Bi_3Se_4$ and $Bi_4Se_3$ hexagonal phases with small bismuth spheres on top.

*Table S2* EDS (*XRF) analysis of the samples pressed at 730°C presented in the main text. The composition of the samples is shifted towards a Bi rich/Se poor composition when pressed in graphite die.

| Elements | Pressed in C-die (%) | Flash from C die *(%) | Pressed in Si₃N₄ -die (%) |
|---|---|---|---|
| O | 49.19 | - | 48.06 |
| Se | 16.01 | 21.50 | 17.72 |
| Bi | 34.80 | 78.50 | 34.21 |
| Bi/Se | 2.17 | 3.65 | 1.93 |

*XRF analysis

### 3. Stability problems with temperature cycling

The data presented in Figure S2 are from the first measurement of pristine samples at 300 K and 573 K. Their preparation route did not follow any recommendation for better stability given in the main text. Figures S5a,b shows an example of the data obtained for the same $Bi_2O_2Se$ in the temperature range 300-773 K and the effect of cycling on the electrical conductivity of the sample. The relative changes in electrical conductivity are much larger than the changes in Seebeck coefficient. This is because electrical conductivity is a transport (current) property, while Seebeck coefficient is a potential measurement. This behavior difference between electrical conductivity and Seebeck points to the role of foreign phases at grain boundaries.

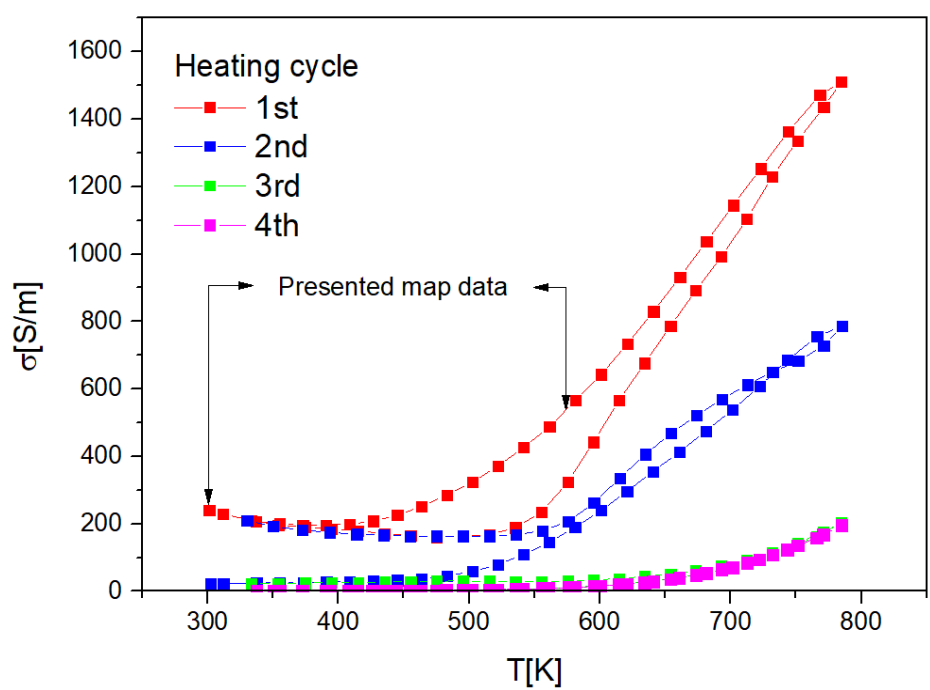

**Figure S5a**. Electrical conductivity as a function of temperature and heating cycles of a $Bi_2O_2Se$ synthesized at 860 °C for 10 days. The sample was pressed from unsieved powder in $Si_3N_4$ die at 730 °C

for 1.5h. The sample thus did not follow any synthesis recommendation for better stability given in the main text

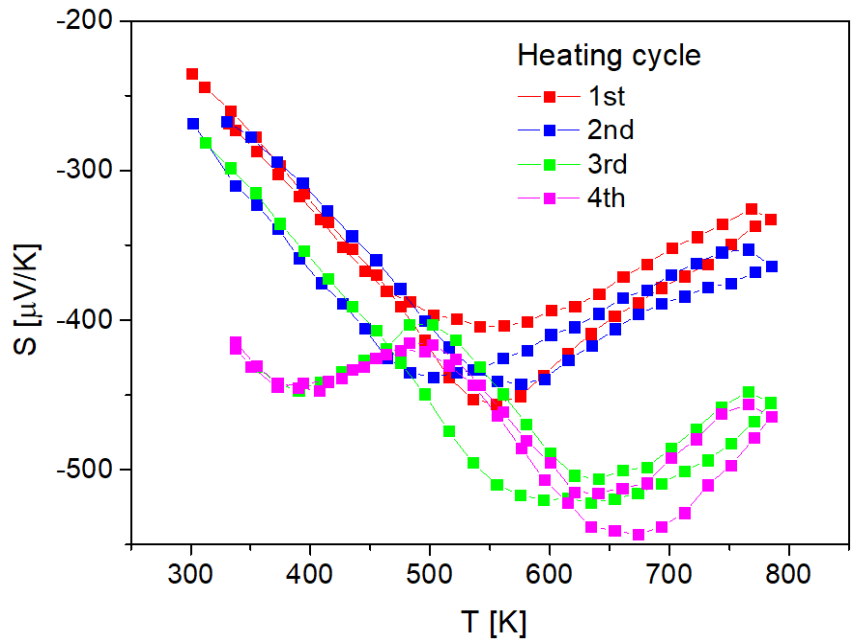

**Figure S5b**. Seebeck coefficient as a function of temperature and heating cycles of a $Bi_2O_2Se$ synthesized at 860 °C for 10 days. The sample was pressed from unsieved powder in $Si_3N_4$ die at 730 °C for 1.5h. The sample thus did not follow any synthesis recommendation for better stability given in the main text

To further illustrate the broad variability in the behaviour of polycrystalline $Bi_2O_2Se$, we prepared multiple samples using various processing techniques. Phase purity of all samples was confirmed by powder X-ray diffraction (XRD), and their transport properties were evaluated after hot pressing.

All samples were synthesized via the low-temperature method, followed by a high-temperature crystal growth step in a temperature gradient, described in the main text. After synthesis, the resulting ingots were sieved to obtain defined particle size fractions for hot pressing. However, it was observed that dry sieving of the as-grown material frequently yielded fractions containing agglomerates and residual fine particles, which were not effectively removed by mechanical separation alone.

**Sample HP730(100-250)** was prepared from as sieved 100–250 μm fraction pressed in $Si_3N_4$ die at 730°C and 70 MPa for 1.5h.

**Sample HP730(100-250 toluene)** was prepared from 100–250 μm fraction cleaned using an ultrasonic ring (Sonoplus, SH 213 horn with VS70T sonotrode, Bandelin, Germany) for five cycles (30 s on / 30 s off) at 100 % amplitude in toluene. The aim of this procedure was to address limitations associated with residual fine particles and agglomerates, The resulting powder was decanted, rinsed in toluene, vacuum-dried at $10^{-1}$ Pa for 5 minutes and hot pressed in a $Si_3N_4$ die at 730 °C and 70 MPa for 1.5 h. This step effectively removed fine particles from powder. Toluene was selected for its ease of evaporation and potential to influence grain boundary chemistry through carbon residues.

**Sample HP730(100-250 water)** used the as sieved 100–250 μm fraction milled in a MM500 Nano oscillating mill at 25 Hz for 1 hour in water. This approach aimed to investigate the effects of particle size reduction and exposure to aqueous media.

Although all samples exhibited phase purity by XRD, their transport properties and stability differed significantly (Figure S6). Notably, both the toluene-cleaned coarse particle sample and the water-milled fine particle sample demonstrated stability under repeated thermal cycling, indicating that appropriate processing can mitigate degradation mechanisms otherwise common in $Bi_2O_2Se$ polycrystals.

*Table S3. Effect of thermal cycling on hot-pressed powders at 730°C for 1.5h, with various powder treatments before pressing. The table shows the difference between initial values of Seebeck and electrical conductivity and their change after two measurement cycles*

| HP730 (100-250) | As prepared (T = 350 K) | | After 2 cycles (T = 350 K) | | Difference | |
|---|---|---|---|---|---|---|
| | σ (S m$^{-1}$) | S (μV K$^{-1}$) | σ (S m$^{-1}$) | S (μV K$^{-1}$) | **Δσ (%)** | **ΔS (%)** |
| pure | 637.1 | -433.6 | 352.3 | -463.1 | **-44.7** | **6.8** |
| toluene | 1554.9 | -262.9 | 1541.7 | -263.0 | **-0.8** | **0.0** |
| water | 9.2 | -482.2 | 4.9 | -481.9 | **2.2** | **-0.1** |

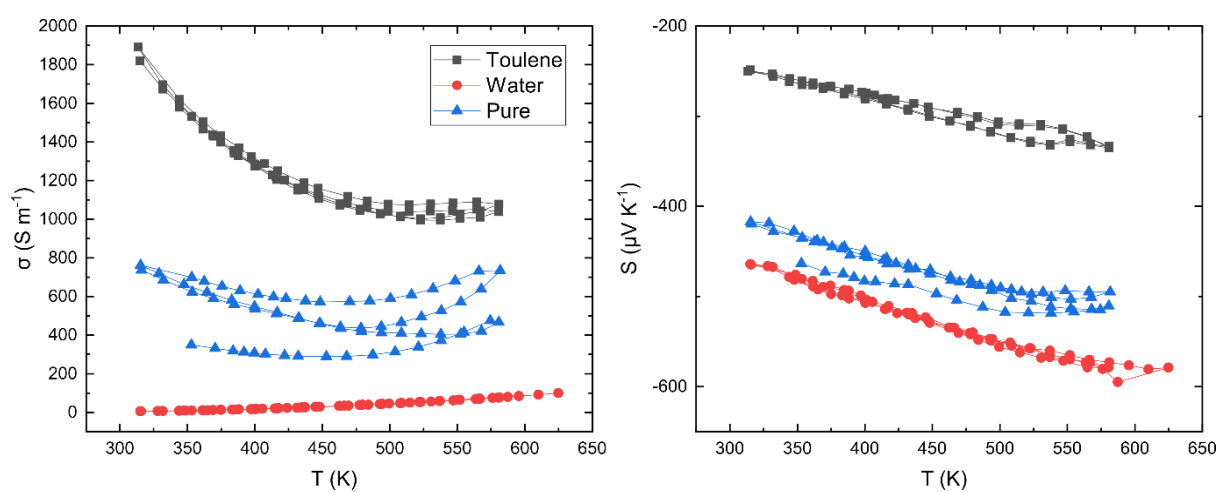

**Figure S6**. Comparison of electrical conductivity and Seebeck coefficient as a function of temperature and heating cycles of a $Bi_2O_2Se$ samples prepared using hot pressing in $Si_3N_4$ die. Sample sonificated in toluene showed increased conductivity and better cycling stability than pure as sieved sample. Water milled sample also showed good stability but had significant decrease in electrical conductivity.

## Pressing temperature effect in SPS

To assess effect of increasing temperature is SPS, we prepared three samples from 100-250um fraction at temperatures of 580, 650 and 730°C. While we can see that increasing pressing temperature leads to higher initial conductivity, instability of samples are still present. This may also be caused by their porosity given low density, as shown in table S4.

*Table S4* Densities of SPS samples from the 100–250 μm fraction, pressed at 580 °C, 650 °C, and 730 °C.

| T (°C) | ρ (g.cm$^{-1}$) | ρ (teor. %) |
|---|---|---|
| 580 | 8.48 | **89.1** |
| 650 | 8.78 | **92.3** |
| 730 | 8.96 | **94.2** |

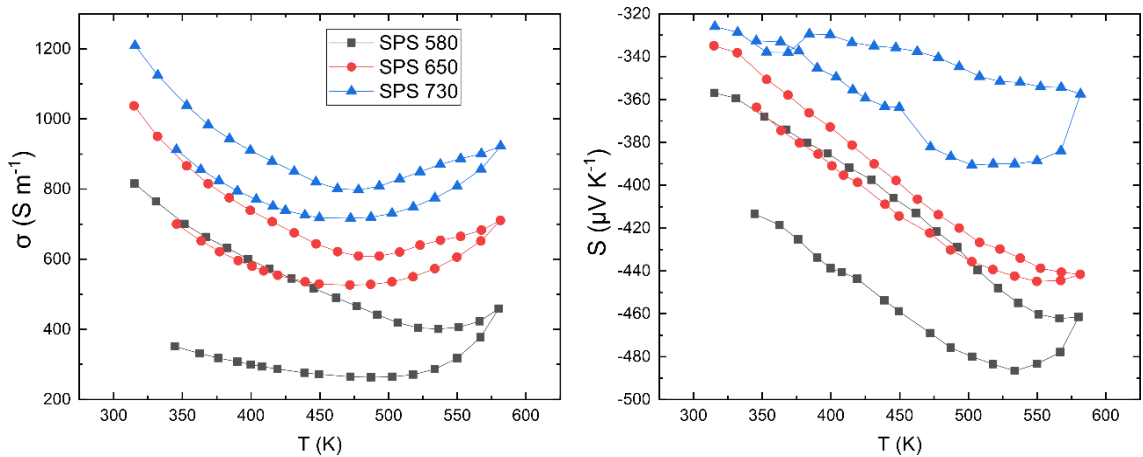

**Figure S7**. Comparison of electrical conductivity and Seebeck coefficient as a function of temperature and one heating cycle of a Bi$_2$O$_2$Se samples prepared using spark plasma sintering in graphite die at various temperatures of pressing. Graph shows higher pressing temperatures is leading to higher conductivity, similarly to hot pressed samples in graphite die.

## Inhomogeneity in SPS samples

Since the SPS samples were 20 mm diameter, we were able to cut multiple bar samples for transport properties measurement. Figure S8 shows two samples with position of cut. As the samples were - technically said - identical, we expected difference in values to be close to measurement precision. However, we observed that even in one pressed sample, high variation of properties can be present. One cause could be inhomogeneity of temperature distribution in SPS during fast temperature ramps, but this assumption would need further study to correctly assess these effects.

*Table S5.* Effect of thermal cycling on SPS powders at 730°C for 1.5h, with various powder treatments before pressing. The table shows the difference between initial values of Seebeck and electrical conductivity and their change after two measurement cycles

| Sample | As prepared (T = 350 K) | | After 2 cycles (T = 350 K) | | Difference | |
|---|---|---|---|---|---|---|
| | σ (S m$^{-1}$) | S (µV K$^{-1}$) | σ (S m$^{-1}$) | S (µV K$^{-1}$) | Δσ (%) | ΔS (%) |
| 1 | 744.5 | -360.5 | 395.3 | -388.2 | -46.9 | 7.7 |
| 2 | 1054.1 | -334.5 | 800.8 | -333.8 | -24.0 | -0.2 |

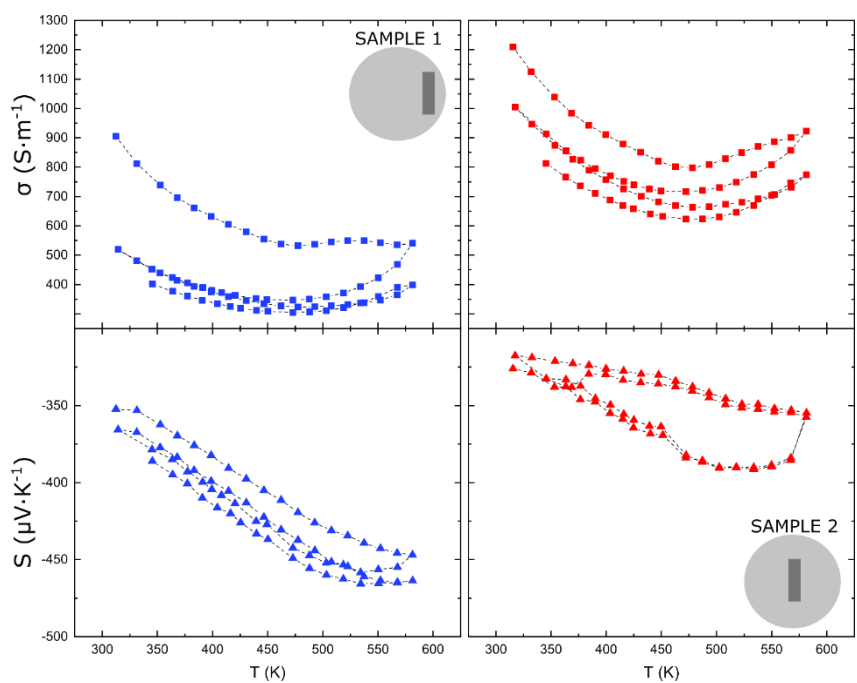

**Figure S8** Comparison of two samples cut from various locations of same pressed pellet of Bi$_2$O$_2$Se. Powder fraction was 100 - 250 µm and pressing was done using method described in main text. Samples were cut without any solvent and dry polished.

These results indicate that all variables noted in main text (grain size, quality of each grain, grain surface, orientation of grains to each other) play crucial roles in final properties of material. Low density of pressed samples (< 95 %) leads to open porosity, allowing faster degradation of surfaces. With open porosity, even polishing or Archimedes density measurement in organic solvent could cause degradation of sample during measurement, as organic solvents adhere strongly to surface of Bi2O2Se and can react at elevated temperatures used during measurement. Also, while XRD is widely recognized method for phase purity identification of material, in case of grain surface-dependent properties, it may hide important information.

### 4. Difficulties in performing doping studies

Table S6 shows that most doped polycrystalline materials contain foreign phases, especially when doping elements substitute Bi. We reproduced some of these studies and found foreign phases. Due to the best match of the crystal/ion radii, Ce and La are likely to slightly dissolve in Bi$_2$O$_2$Se, as evidenced by the clear validity of Vegard's law[4,5]. Generally, however, the solubility of the doping metals in the Bi sublattice is very low, indicating the extreme stability of the $Bi_2O_2$ layer.

Importantly, the formation of foreign phases is largely associated with changes in the stoichiometry of the $Bi_2O_2Se$ matrix – in particular, when the doping metal prefers oxidation state different from Bi, i.e. $Bi^{+3}$. This can be most easily seen on some examples like Ti (see main text) or Ta, Mn, W etc. Assume that we plan to substitute Ta for Bi in the form of $Bi_{2-x}Ta_xO_2Se$. When $Ta_2O_5$ segregates as a foreign phase instead of substituting in $Bi_2O_2$ layer, we are left with an O-poor Se-rich composition of matrix:

$$Bi_{2-x}Ta_xO_2Se \quad \rightarrow \quad Bi_{2-x}O_{2-2.5x}Se + xTa_2O_5. \qquad (1)$$

Such a formation of FP shows that the matrix stoichiometry may change, and hence, the character and concentration of native defects and other FPs (see the section 3.2 in the main text). On the other hand, the FPs can also directly transfer charge to the $Bi_2O_2Se$ matrix, which is essentially a modulation doping. This is often facilitated by the 2D character of both the matrix ($Bi_2O_2Se$) and foreign phase (e.g. BiSe, $Bi_4Se_3$, $Bi_2Se_3$ [6]. In addition to the very poor reproducibility of the literature and our own data, we particularly emphasize the fact that pure polycrystalline $Bi_2O_2Se$ is very difficult to obtain, and its TE properties are very poor. Table S7 summarizes the synthesis steps used in the published studies. It is intended for comparison with the synthesis steps proposed in the present work.

The change in temperature during cycling shifts the chemical equilibrium and may subsequently lead to the reaction of foreign phases with $Bi_2O_2Se$ matrix (see the reactions shown in main text and the list of reactions below). The eventual change in composition (type and number of foreign phases and matrix stoichiometry) in the solid will essentially depend on the temperature and its rate of change in terms of reaction kinetics. The same can be true for undoped single crystals, which have a wide range of transport properties (Table S8).

Table S6. Properties of undoped samples used as standard for doped samples in doping studies.

| Doped Element | ZT | S | σ | Reported Solubility Limit | Foreign Phases | Ref. |
|---|---|---|---|---|---|---|
| | (-) | (μV/K) | (S/cm) | (at. %) | | |
| Ti | 0.47 (772 K) | -130 (300 K)<br>-240 (772 K) | 124 (300 K)<br>75 (772 K) | 0.5 | $TiO_2$ | [7] |
| W | 0.48 (773 K) | -130 (300 K)<br>-250 (773 K) | 175 (300 K)<br>70 (773 K) | 0.5 | $Bi_2WO_6$ | [8] |
| Nb | 0.06 (823 K) | -600 (323 K)<br>-450 (873 K) | 2 (323 K)<br>5 (873 K) | 2 | $Nb_2O_5$ | [10] |
| Ag | 0.06 (673 K) | -360 (300 K)<br>-520 (673 K) | 0.1 (300 K)<br>1.5 (673 K) | insoluble | $Ag_2Se$;<br>$Bi_2O_3$; Ag | [11] |
| Ta | 0.1 (773 K) | -360 (300 K)<br>-400 (873 K) | 0.1 (300 K)<br>8 (873 K) | 3* | $Ta_2O_5$ | [12] |
| Ge | 0.18 (780 K) | -170 (300 K)<br>-276 (780 K) | 83 (300 K)<br>25 (780 K) | 2.5 | $Bi_4(GeO_4)_3$;<br>BiSe | [13] |
| Ge | 0.13 (823 K) | -269 (300 K)<br>-370 (823 K) | 0.03 (300 K)<br>8 (823 K) | 3* | $Bi_4(GeO_4)_3$ | [14] |
| La | 0.06 (823 K) | -440 (300 K)<br>-420 (823 K) | 0.1 (300 K)<br>8 (823 K) | 2 | $La_2O_3$ | [4] |
| Sb | 0.33 (773 K) | -225 (323 K)<br>-340 (773 K) | 110 (323 K)<br>70 (773 K) | 1 | $BiSbO_4$;<br>$Bi_4Se_3$ | [15] |
| Ce | 0.19 (773 K) | -189 (300 K)<br>-299 (773 K) | 0.1 (300 K)<br>20.4 (773 K) | 2.25 | $CeO_2$; $Bi_2O_3$ | [5] |
| Sn | 0.05 (773 K) | -200 (300 K)<br>-450 (773 K) | 0.1 (300 K)<br>2.3 (773 K) | 5* | - | [16] |
| Zr | 0.11 (773 K) | -152 (323 K)<br>-243 (773 K) | 5.14 (323 K)<br>12 (773 K) | 3* | - | [17] |
| Te ($Te_O$) | 0.45 (773 K) | -90 (323 K)<br>-180 (773 K) | 270 (323 K)<br>160 (773 K) | 0.5 | $TeO_2$;<br>$Bi_2O_2Te$;<br>$Bi_3Se_4$ | [18] |
| Te ($Te_{Se}$) | 0.10 (823 K) | -315 (300 K)<br>-420 (823 K) | 0.1 (300 K)<br>7.0 (823 K) | 6* | $Bi_2O_3$ | [18] |
| S ($S_O$) | 0.10 (793 K) | -350 (323 K)<br>-480 (793 K) | 0.13 (323 K)<br>8.0 (793 K) | 0.5 | $Bi_2S_3$;<br>$Bi_2O_{2.33}$ | [19] |
| Cl ($Cl_{Se}$) | 0.08 (823 K) | -240 (300 K)<br>-340 (823 K) | 0.1 (300 K)<br>6.2 (823 K) | 1.5 | $Bi_{12}O_{15}Cl_6$ | [20] |
| I ($I_{Se}$) | 0.21 (790 K) | -232 (300 K)<br>-316 (790 K) | 15.9 (300 K)<br>22.5 (790 K) | 0.75* | - | [21] |

*max presented concentration in work and dopant is still soluble

Table S7. Synthesis procedure used in studies. *The number of references refers to the main article.*

| Ref | $\sigma_{300K}$ (S.m$^{-1}$) | Raw materials | Synthesis |
|---|---|---|---|
| [35] | 3 | Bi$_2$O$_3$ 4N; Bi 4N; Se 4N | Vac BM 4hrs, CP -> QA 573 K/ 6 hrs -> 773 K/ 12 hrs, Vac BM 4hrs, SPS 903 K/ 5 min/ 50MPa |
| [25] | 10 | NA | Vac BM 4hrs -> CP -> QA 573/ 6 hrs->773/ 12, Vac BM 4 hrs SPS 823 K/ 5 min/ 50MPa, BM 6hrs |
| [17] | 10 | Bi$_2$O$_3$ 4N; Bi 2N5; Se 2N5 | Manual mixing and crushing, CP -> QA 573 K/ 6hrs ->773 K/ 12 hrs, SPS 903 K/ 5 min/ 50MPa |
| [14] | 10 | Bi$_2$O$_3$ 4N; Bi 4N; Se 4N | Vac BM 4 hrs -> CP -> QA 573 K/ 6 hrs -> 773 K/ 12 hrs, Vac BM 4 hrs, SPS 903 K/ 5 min/ 50MPa, BM 6 hrs |
| [13] | 13 | NA | solid state reaction under argon 873 K/ 24 hrs, SPS 873 K/ 5 min/ 50 MPa |
| [18] | 200 | Bi$_2$O$_3$ 4N; Bi 4N; Se 4N | BM 4hrs, CP -> QA 573 K/ 6 hrs -> 773 K / 12 hrs, HP 853 K/ 1hrs/ 65 MPa |
| [12] | 514 | Bi$_2$O$_3$ 5N; Bi 4N; Se 5N | High-energy BM on air, CP -> QA 673K/ 2hrs |
| [11] | 1590 | Bi$_2$O$_3$ 4N; Bi 4N; Se 5N | QA 873 K/ 24 hrs -> BM crushing, SPS 873 K/ 5 min/ 50MPa |
| [24] | 8300 | Bi$_2$O$_3$ 5N; Bi 5N; Se 5N | QA 973 K/ 10 days, Milling in hexane. HP 853 K/ 1 hr/ 70 MPa |
| [10] | 11000 | NA | QA 873 K/ 24 hrs, shear-exfoliation in water -> drying on air, SPS 5 min/ 50 MPa |
| [27] | 12000 | Bi$_2$O$_3$ 2N; Bi 4N; Se 4N | Manual mixing and crushing, QA 873 K/ 24hrs, shear-exfoliation in water, SPS 873 K/ 5 min/ 50MPa |
| [16] | 12400 | Bi$_2$O$_3$ 2N; Bi 4N; Se 4N | Wet grounding, Ar-QA 873K/ 24hrs, shear-exfoliation in water -> drying on air, SPS 5 min/ 50 MPa |
| [19] | 17500 | NA | QA 873K/ 24hrs, shear-exfoliation in water, -> drying on air on air, SPS 5 min/ 50 MPa |
| [29] | 27000 | Bi$_2$O$_3$ 2N; Bi 4N; Se 4N | Manual mixing and crushing, QA 873K/ 24hrs, shear-exfoliation in water -> drying on air, SPS 873 K/ 5 min/ 50MPa |

*BM – ball milling; QA – sealed in evacuated quartz ampoules; SPS – spark plasma sintering; HP – hot-pressing; CP – cold-pressing;*

Table S8 lists a range of Bi$_2$O$_2$Se single crystals from various sources, highlighting that this material can exhibit problematic behavior even in single-crystalline form. In general, precise control over doping induced by native defects or non-stoichiometry remains challenging. Notably, even "seemingly perfect" single crystals may display wide variability in properties and, in many cases, significant deviations from stoichiometry, as documented, for example, in Ref. [24].

Note that in terms of single crystals, the problem with decarbonation is most likely irrelevant to the preparation of single crystal nanoparticles[22,23] prepared by chemical vapor transport. Conversely, it may be problematic for single crystals prepared from melt by the Bridgman method[24–26] or by chemical transport in evacuated closed SiO$_2$ ampoules[27]. The variety of properties of single crystals indicates some inconsistency.

**Table S8.** Properties of undoped Bi$_2$O$_2$Se single crystals.

| Single crystals Ref. No. | S$_{(300\ K)}$ (µV·K$^{-1}$) | σ$_{(300\ K)}$ (S·cm$^{-1}$) | σ (S·cm$^{-1}$) | n$_{300K}$ (10$^{18}$·cm$^{-3}$) | N$_{(2K)}$ (10$^{18}$·cm$^{-3}$) | µ$_{H(300\ K)}$ (cm$^2$·V$^{-1}$·s$^{-1}$) | µ$_{H(2K)}$ (cm$^2$·V$^{-1}$·s$^{-1}$) |
|---|---|---|---|---|---|---|---|
| Mao et al.[28] | NA | 6.5 | 250 | 4.8 | 2.1 | 10 | 800 |
| Tong et al.[25] | NA | 2000 | 157 000 | 23.0 | 19.0 | 900 | 42 000 |
| Lv et al.[29] | NA | 666 | 300 000 | 15.0 | 11.0 | 300 | 220 400 |
| Chen et al.[30] | NA | NA | NA | 8.5 | 6.6 | 370 | 280 000 |
| Wang et al.[27] | NA | NA | NA | NA | 5.0 | NA | 320 000 |
| Drašar et al.[31] | -144 | 330 | NA | 7.4 | NA | 307 | NA |
| Li et al. (Se poor)[32] | NA | 2222 | 32000 | 400 | 180 | 50 | 4500 |
| Li et al. (Se rich)[32] | NA | 4000 | 40000 | 1400 | 610 | 25 | 280 |

The carrier mobility of nominally undoped samples reaches up to 470000 and 280000 cm$^2$V$^{-1}$s$^{-1}$ at 2 K[30,33], which not only exceeds theoretical predictions (≈100000 cm$^2$V$^{-1}$s$^{-1}$, for a carrier concentration of N=10$^{16}$ cm$^{-3}$)[34], but is incomparably higher than the ≈100 - 5000 cm$^2$V$^{-1}$s$^-$1 published in[32]. In addition, the deviation in stoichiometry (Bi/Se=1.4 -2.3) for seemingly perfect single crystals is particularly worrying[32]. It indicates both the presence of "invisible" foreign phases (Bi-Se or Bi$_2$SeO$_5$ based) and a possible large concentration of native defects. In addition, the reported effective electron mass ranges from 0.03*m$_e$[32] to 0.23*m$_e$[27], which is not conclusive. Also, its low-temperature T-square dependence of the resistivity remains elusive in the context of today's models[27]. Thus, a pure "standard" single crystal does not yet exist. Therefore, although we present a brief characterization of our "average" single crystal in the SI (Figure S15), it cannot be considered representative. This material shows metal-insulator transition concentration of free carriers as low as $n \approx 10^{15}$ cm$^{-3}$ [35]. Doped samples with a much higher carrier concentration should therefore be metallic. This implies that any feature of " semiconducting" behavior observed in doped materials is due to grain boundaries/foreign phases. This also explains the lacking single-crystalline "standard". Unfortunately, these two lines (basically grain boundaries properties and grain volume properties) is difficult to tell apart. In this regard, understanding the behavior of polycrystalline $Bi_2O_2Se$ is highly problematic.

## 5. List of reactions that can occur during measurement or compaction

In this section we present other equilibrium reactions that can take place during temperature changes in $Bi_2O_2Se$ polycrystals. This list complements the reactions mentioned in the main text. For energetic reasons, these reactions will preferentially take place at the grain surface, i.e., at grain boundaries in polycrystals.

If only $CO_2$ is present in any way then the second reaction comes into play, which runs clearly to the right

$$3\ Bi_2O_2Se + 2\ CO_2 \leftrightarrow 2\ Bi_2CO_5 + Bi_2Se_3 \quad (\Delta H=-97\ kJ\ (-1.0\ eV)/reaction).$$

When both $CO_2$ and $O_2$ is present in any way then the second reaction comes into play

$$8\ Bi_2O_2Se + 3O_2 + 4CO_2 \leftrightarrow 2Bi_2SeO_5 + 2Bi_2Se_3 + 4Bi_2CO_5\ (\Delta H=-1331\ kJ\ (-13.8\ eV)/reaction).$$

Other reactions are for example

$$2\ Bi_2O_2Se + 2CO_2 + SeO_2 \leftrightarrow 3Se + 2Bi_2CO_5 \quad (\Delta H=-92\ kJ\ (-0.96\ eV)/reaction).$$

This reaction can be followed by

$$4\ Bi_2O_2Se + SeO_2 + 3Se \leftrightarrow 2Bi_2SeO_5 + 2Bi_2Se_3 \quad (\Delta H=-143\ kJ\ (-1.49\ eV)/reaction).$$

The reaction enthalpies are relatively small (<0.4 eV per unit of formula $Bi_2O_2Se$), suggesting that the equilibrium point of these reactions will shift significantly with temperature and composition[36,37].

## 6. Difficulties in foreign phase evidence in doped materials

In this section, we show that using PXRD as a first choice when searching for extraneous phases can be tricky. Using the example of Mn doping, we will show that the foreign phase ($Mn_3O_4$) is clearly present in SEM - BSE images but invisible to PXRD - $Bi_{1.9}Mn_{0.1}O_2Se$ sample (Figures S9 and S10). This is probably due to the formation of very thin precipitates in the matrix of 2D nature. It is visible to PXRD for $Bi_{1.8}Mn_{0.2}O_2Se$ sample (Figures S11 and S12).

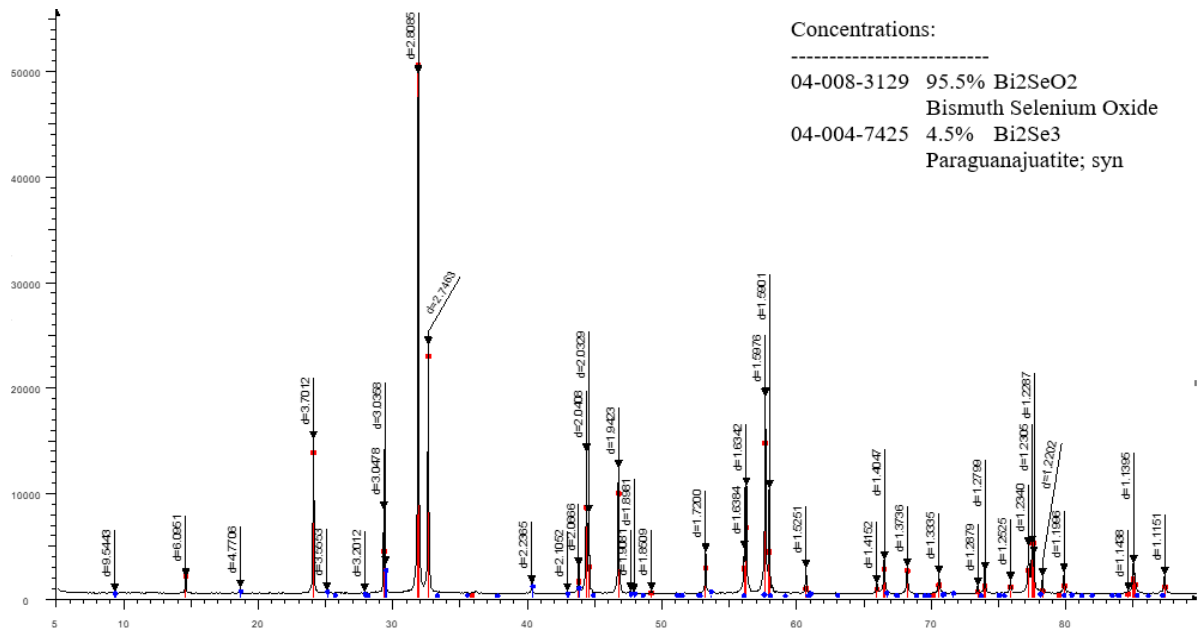

**Figure S9.** PXRD diffractogram of $Bi_{1.9}Mn_{0.1}O_2Se$ sample, only $Bi_2O_2Se$ (red peaks) and $Bi_2Se_3$ (blue peaks) phases are present. No Mn-based oxides detectable.

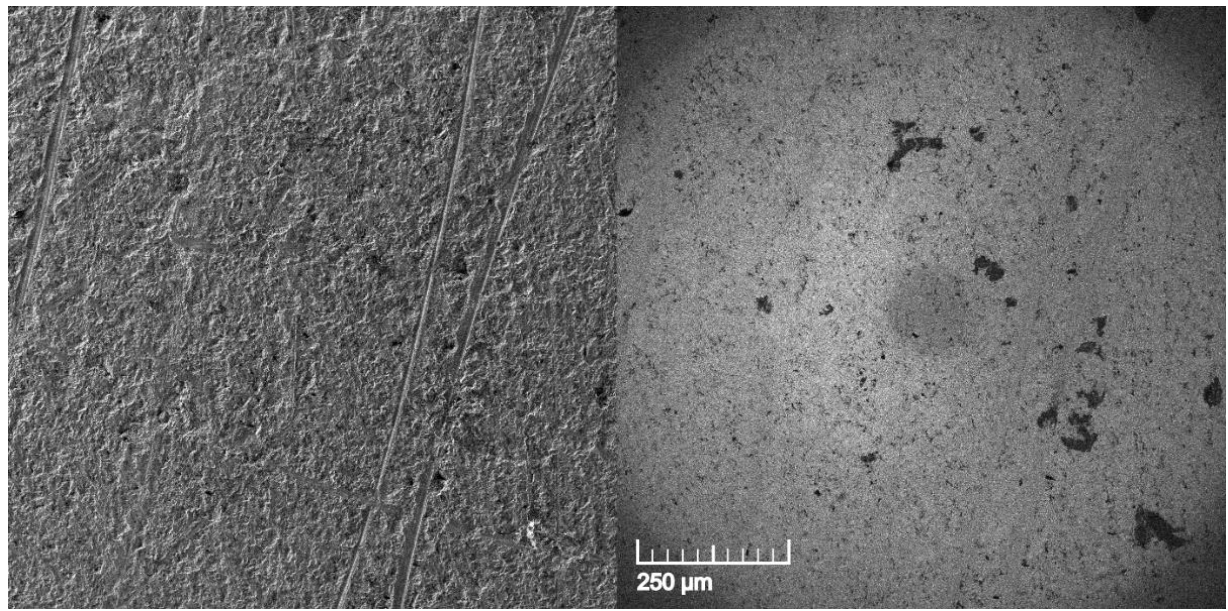

**Figure S10.** SEM and BSE images of $Bi_{1.9}Mn_{0.1}O_2Se$ sample; black areas in the BSE image represent lighter elements – in this case $Mn_3O_4$. This contrasts with XRD in Fig. S9 where $Mn_3O_4$ is undetectable.

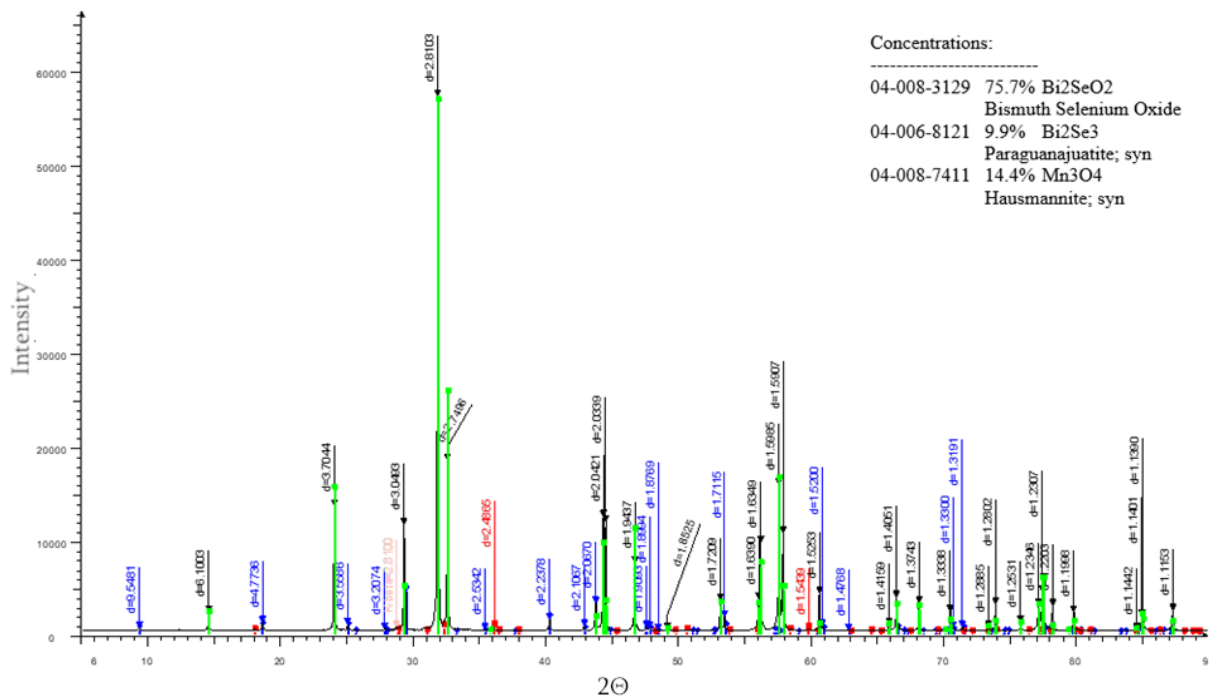

**Figure S11.** XRD diffractogram of $Bi_{1.8}Mn_{0.2}O_2Se$ sample, $Bi_2O_2Se$ (green peaks) and $Bi_2Se_3$ (blue peaks) and $Mn_3O_4$ (red peaks) phases are present.

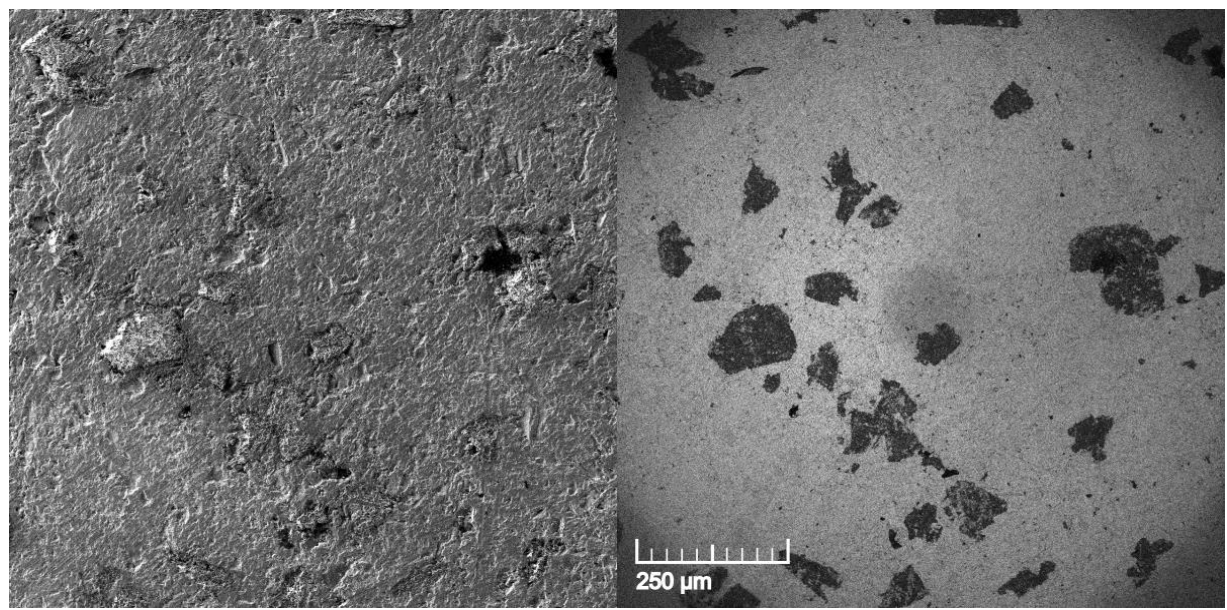

**Figure S12.** SEM and BSE images of $Bi_{1.8}Mn_{0.2}O_2Se$ sample, black areas in the BSE image represent lighter elements – in this case $Mn_3O_4$.

**Table S9.** EDS analysis of "dark spots" $Mn_3O_4$.

| | Composition of dark spots (at%) |
|---|---|
| O | 60.44 |
| Mn | 37.55 |
| Se | 0.76 |
| Bi | 1.21 |

The EDS analysis in Table S10 shows an inhomogeneous distribution of Mn in the $Bi_2O_2Se$ matrix.

*Table S10. EDS analysis of $Bi_{1.9}Mn_{0.1}O_2Se$ and $Bi_{1.8}Mn_{0.2}O_2Se$ samples (average of 8 measurements each).*

|    | $Bi_{1.9}Mn_{0.1}O_2Se$ | $Bi_{1.8}Mn_{0.2}O_2Se$ |
|----|-------------------------|-------------------------|
| O  | 62.38                   | 56.31                   |
| Mn | 0.53                    | 6.24                    |
| Se | 11.65                   | 12.55                   |
| Bi | 25.45                   | 24.91                   |

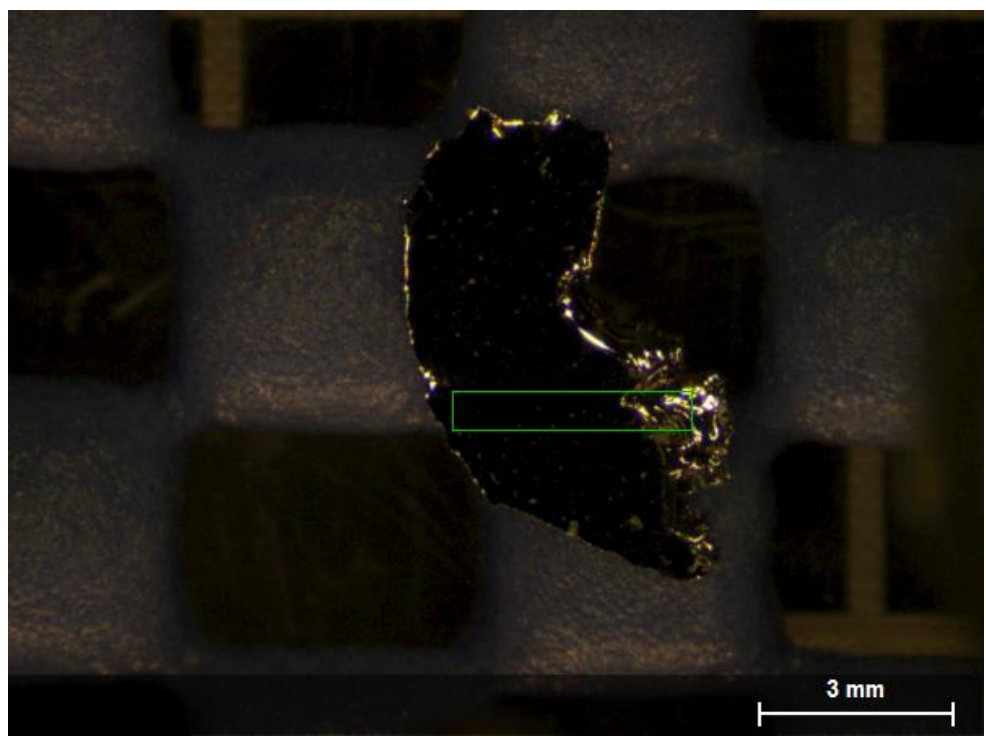

**Figure S13.** $Bi_{1.9}Mn_{0.1}O_2Se$ single crystal and green rectangle indicating measurement area of XRF analysis.

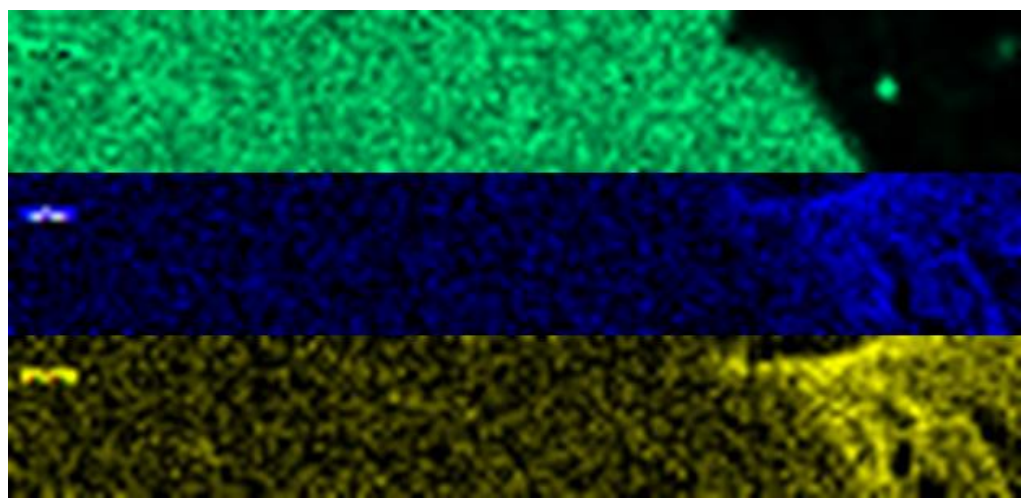

**Figure S14.** XRF measurement of $Bi_{1.9}Mn_{0.1}O_2Se$ single crystal, green shows Mn distribution, blue shows Se distribution and yellow shows Bi distribution (O can't be measured by this method).

**Table S11.** Composition of highlighted area from Fig. S13 of $Bi_{1.9}Mn_{0.1}O_2Se$ single crystal according to XRF analysis.

|    | Composition (At%) |
|----|-------------------|
| Mn | 2.5               |
| Se | 26.78             |
| Bi | 70.72             |

## 7. Properties of $Bi_2O_2Se$ single crystals

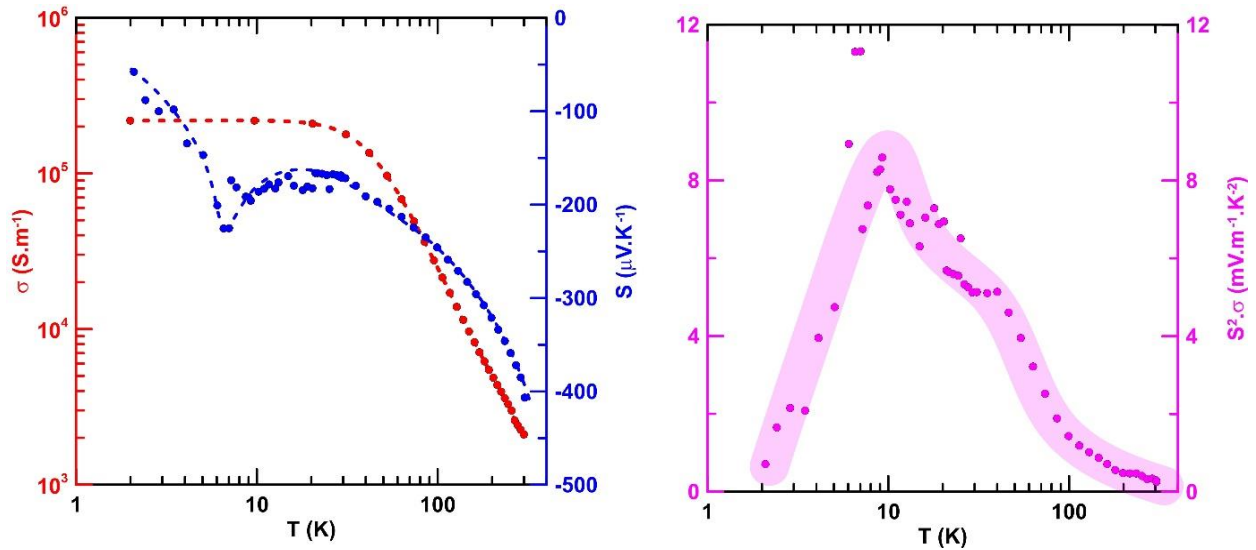

**Figure S15.** Thermoelectric properties of the $Bi_2O_2Se$ single crystal obtained in this study. Notice that the room temperature electrical conductivity is three times higher than electrical conductivity of corresponding polycrystal pressed in $Si_3N_4$ die presented in main text (Figure 4). This can be attributed either to grain boundaries scattering or to the randomness of properties of individual crystals within the batch (ref. [27]). The experiment was performed on PPMS apparatus by Quantum Design.

## 8. Thermoelectric figure of merit

Although the objective of this paper is to establish a standard methodology for the preparation of a "standard" undoped sample, we provide the reader with the measurement of the thermoelectric figure of merit $ZT = \sigma S^2/\kappa$. Due to the preferential orientation the thermal conductivity $\kappa$ was measured on rectangular samples (10 x 10 x 2 mm$^3$) cut along the major axes of a cylinder 10 mm high and 12 mm in diameter. Thus, the orientation corresponded to the measurements of the Seebeck coefficient $S$ and electrical conductivity $\sigma$. First, the thermal diffusivity $k$ was measured. The thermal conductivity κ was then calculated using the relationship $\kappa = kc_P\rho$, where $\rho$ is the experimental density and $c_P$ is the heat capacity. Inconel was used as the heat capacity standard. The experimental values of the heat capacity $c_P$ were close (0.220 - 0.255 Jg$^{-1}$K$^{-1}$) to the Dulong-Petit value $c_V$ = 0.236 Jg$^{-1}$K$^{-1}$ at T=300 K. Thermal diffusivity and heat capacity were measured with LFA 457 (Netzsch, Germany).

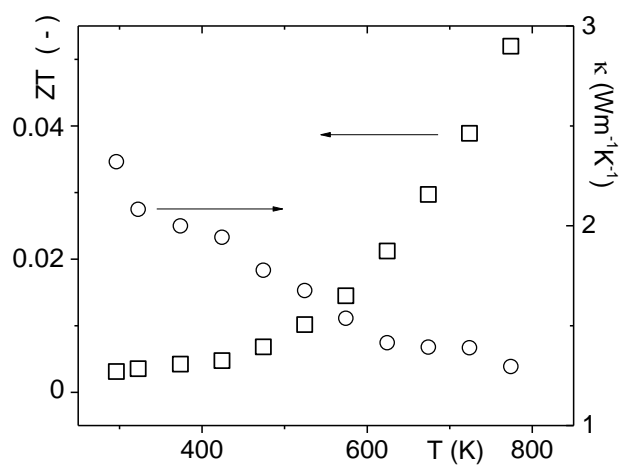

**Figure S16.** Thermoelectric figure of merit ZT= f(T) of $Bi_2O_2Se$ sample prepared according to the 5 recommendations in the main text - pressed in $Si_3N_4$ die at 730°C for 3 hours; sieved and purified coarse grained material, fraction 35 - 340 μm. Note that the ZT for "pure" undoped material is incomparably lower than for doped materials presented in Table 6. The thermal conductivity in the in-plane direction is 1.5 times higher than in the out-of-plane direction.